\documentclass[11pt]{article}

\usepackage[english]{babel}
\usepackage{amsmath,amsthm}

\usepackage{latexsym} 
\usepackage{amsfonts}
\usepackage{amsmath}
\usepackage{amsthm}
\usepackage{amssymb}
\renewcommand{\theequation}{\thesection.\arabic{equation}}
\newcounter{subequation}[equation]
\makeatletter

\newcommand{\p}{^{\prime}}

\expandafter\let\expandafter\reset@font\csname reset@font\endcsname

\def\subeqnarray{\arraycolsep1pt
	\def\@eqnnum\stepcounter##1{\stepcounter{subequation}%
		{\reset@font\rm(\theequation\alph{subequation})}}
	\jot5mm     \eqnarray}

\makeatother

\def\be{\begin{equation}}
\def\ee{\end{equation}}

\def\lb{\label}

\def\bea{\begin{eqnarray}}
\def\eea{\end{eqnarray}}
\def\ba{\begin{array}}
	\def\ea{\end{array}}
\def\dd{\partial}

\def\half{\frac{1}{2}}

\def\one#1{#1^{\raise5pt\hbox{$\scriptstyle\!\!\!\!1$}}\,{}}
\def\two#1{#1^{\raise5pt\hbox{$\scriptstyle\!\!\!\!2$}}\,{}}

\def\tilde{\widetilde}
\def\II{\hbox{{1}\kern-.25em\hbox{l}}}
\def\a{\alpha}
\def\b{\beta}

\def\d{\delta}
\def\e{\varepsilon}
\def\qed{\rule{5pt}{5pt}}
\makeatletter
\def\binrel@#1{\begingroup
	\setboxz@h{\thinmuskip0mu
		\medmuskip\m@ne mu\thickmuskip\@ne mu
		\setbox\tw@\hbox{$#1\m@th$}\kern-\wd\tw@
		${}#1{}\m@th$}%
	\edef\@tempa{\endgroup\let\noexpand\binrel@@
		\ifdim\wdz@<\z@ \mathbin
		\else\ifdim\wdz@>\z@ \mathrel
		\else \relax\fi\fi}%
	\@tempa
}
\let\binrel@@\relax
\def\overset#1#2{\binrel@{#2}%
	\binrel@@{\mathop{\kern\z@#2}\limits^{#1}}}
\def\underset#1#2{\binrel@{#2}%
	\binrel@@{\mathop{\kern\z@#2}\limits_{#1}}}
\makeatother
\newfont{\bbd}{msbm10 scaled\magstep1}

\newtheorem{proposition}{Proposition}

\begin{document}

\begin{center}

     {\LARGE {Orthogonal and symplectic Yangians: \\
linear and quadratic evaluations}}

{\large \sf
D. Karakhanyan$^{a}$\footnote{\sc e-mail: karakhan@yerphi.am  },
R. Kirschner$^b$\footnote{\sc e-mail:Roland.Kirschner@itp.uni-leipzig.de}} \\

\vspace{0.5cm}

\begin{itemize}
\item[$^a$]
{\it Yerevan Physics Institute,
 2 Alikhanyan br., 0036 Yerevan, Armenia }
\item[$^b$]
{\it Institut f\"ur Theoretische
Physik, Universit\"at Leipzig, \\
PF 100 920, D-04009 Leipzig, Germany}
\end{itemize}
\end{center}

\vspace{.5cm}
\begin{abstract}
\noindent
Orthogonal or symplectic Yangians are defined by the Yang-Baxter $RLL$
relation involving the fundamental $R$ matrix with $so(n)$ or $sp(2m)$
symmetry. Simple $L$ operators with linear or quadratic dependence on the
spectral parameter exist under restrictive conditions. These conditions are
investigated in general form.
\end{abstract}

\vspace{2cm}

    \renewcommand{\refname}{References.}
    \renewcommand{\thefootnote}{\arabic{footnote}}
    \setcounter{footnote}{0}

\newpage

\section{Yangian symmetries}

Yang-Baxter relations appeared historically in the study of idealized
models of physical problems of particle scattering and thermodynamics.
They define infinite dimensional algebras related to an underlying Lie
algebra $\mathcal{G}$.
The related physical models have an extended symmetry appearing
in a large number of conservation laws beyond the ones related to the Lie
algebra.
These relations are  well known as the  basis of the treatment of
quantum integrable models \cite{FST,TTF,KuSk1,Fad}.
In recent years the range of applications of  the extended symmetries has
been broadened  essentially, in particular to the study of gauge field
theories, and the related methods have attracted increasing interest
\cite{LevPadua,FK,BSt,DHP}.

The formulation of the Yangian algebra of type
$\mathcal{G}$, $\mathcal{Y}(\mathcal{G})$, can be based on the Yang-Baxter
$R$ matrix acting on the tensor product of the fundamental representation
space $V$ of the Lie algebra $\mathcal{G}$ and obeying the
Yang-Baxter relation of the form
    \be\label{ybe}
    R_{b_1b_2}^{a_1a_2}(u)R_{c_1b_3}^{b_1a_3}(u+v)
 R^{b_2b_3}_{c_2c_3}(v)
 =R_{b_2b_3}^{a_2a_3}(v)R_{b_1c_3}^{a_1b_3}(u+v)
 R^{b_1b_2}_{c_1c_2}(u).
    \ee
The generators $ (L^{(k)})^a_b $ of the extended Yangian algebra
$\mathcal{Y}(\mathcal{G})$ appear in the expansion of the $L$ operator
\be
 \lb{defYan}
 L^a_b(u) = \sum_{k=0}^\infty  \frac{(L^{(k)})^a_b}{u^k}  \,, \qquad
 L^{(0)} = I \, ,
 \ee
which satisfies  the Yang-Baxter RLL-relations
 \be
 \label{rll}
    R^{a_1a_2}_{b_1b_2}(u-v)L^{b_1}_{c_1}(u)L^{b_2}_{c_2}(v)=
    L^{a_2}_{b_2}(v)L^{a_1}_{b_1}(u)R^{b_1b_2}_{c_1c_2}(u-v).
    \ee
$L(u)$ is an algebra valued matrix, $L(u)  \in End\; V \otimes
\mathcal{Y}(\mathcal{G})$,
 depending on the spectral parameter $u$.

The Yangian algebra as originally defined by Drinfeld \cite{Drinfeld,Drin}
is obtained from the extended Yangian by factorizing central elements.
The Yangians of orthogonal and symplectic types have been considered in
\cite{AACFR} and their algebraic structure and representation theory have
been considered in \cite{AMR}.
In the case $\mathcal{G}=g\ell(n)$ the center is contained in the quantum
determinant of $L(u)$. In the cases $\mathcal{G}=so(n)$ and $\mathcal{G}=sp(n)$
the center has been analyzed in \cite{AMR}.
Proofs of the  equivalence of the Yangian definition via (\ref{rll}) and the center
factorization to the definitions by Drinfeld are given in  \cite{JLM17,GRW}.

It is well known that the twofold matrix product $L(u) = L^{(1)}(u)
L^{(2)}(u+\delta)$ (here $\delta$ is an arbitrary shift of the
spectral parameter $u$)
obeys (\ref{rll}), if both  factors obey this $RLL$
relation.
The transfer matrix constructed as the trace of the $N$-fold
product of such $L$ operators called monodromy matrix plays a central
role in the treatment of quantum integrable
models. The relevant case is the one where the  factors entering
the monodromy matrix have a simple form related to the underlying Lie algebra.

If the Lie algebra is the general linear, $\mathcal{G}=g\ell(n)$, the simple
form  $L(u)= Iu + G$ can be chosen without restrictions, where $I$ is the
identity matrix and the matrix elements of $G$ are the Lie algebra
generators. However, in the case of the orthogonal or symplectic Lie algebras,
$\mathcal{G}= so(n)$ or $\mathcal{G}= sp(n)$ ($n$ even), this ansatz linear in $u$
works with essential restrictions only. In this situation
the next-to-simplest case of an ansatz quadratic in $u$  is of importance,
\be \lb{Lo2}
L(u) = Iu^2 + u G+ H .
 \ee
In particular the fundamental $R$ matrix appearing in (\ref{ybe})
in the case of the orthogonal or symplectic symmetry
(compare (\ref{rzW}) below) can be written in the form (\ref{Lo2}) and
(\ref{ybe}) can be regarded as a particular case of (\ref{rll}).
The Jordan-Schwinger class of representations,
where the  generators are composed  of a set of Heisenberg
pairs following the  pattern of Quantum mechanical angular momentum
provide an example of (\ref{Lo2}) \cite{Re,IKK15}.

In general, the truncation of the expansion (\ref{defYan}) of $L(u)$ obeying
(\ref{rll}) by imposing $(L^{(k)})^a_b= 0,  k >p$, results in conditions
defining the order $p$ evaluation of the Yangian algebra,
$\mathcal{Y}^{(p)}(\mathcal{G}) $. The analysis of those conditions becomes
complicated with increasing $p$. It is rather involved already for $p=2$.
Because of its physical relevance this case deserves the effort of
a complete analysis in full generality, which is the central point of this
paper.

Starting from the classical results on the fundamental $R$ matrix with
 orthogonal or symplectic symmetry \cite{ZZ,BKWK,Re} and on   $L$
operators \cite{SW,Re}
the linear and quadratic evaluations
of Yangians of such types have been
studied in recent papers
\cite{CDI1,IKK15,FIKK16,KK16}.
Examples of $L$ operators obeying the condition of the linear or the
quadratic evaluation have been considered in these papers. The spinorial
representation as a case of the linear evaluation has been studied in \cite{AMR}.

In the present paper we go beyond particular examples and derive from the
$RLL$ relations
 (\ref{rll}) with the quadratic ansatz (\ref{Lo2}) the
defining relation of the second order evaluation algebra
$\mathcal{Y}^{(2)}(\mathcal{G}) $.
We resolve the structure of the terms in (\ref{Lo2}) and express them by the
matrices of independent generators $\bar G$ and $\bar H$. We formulate the
algebra relations of $\mathcal{Y}^{(2)}(\mathcal{G})$
as expressions in products of the
latter algebra valued matrices $\bar G$ and $\bar H$.

The constraints implied by the quadratic ansatz are extracted from the
$RLL$ relation in a concise tensor product formulation \cite{FRT},
avoiding an accumulation of indices.  In sect. 3 the simpler case
of the linear evaluation  $\mathcal{Y}^{(1)}(\mathcal{G})$  is reconsidered.
We illustrate in this case how the tensor product formulation is used to
formulate the constraints and also to rewrite them in convenient forms  
in order to understand their meaning.
There  we have only one extra condition on the
matrix of Lie algebra generators. We add some results on the
spinorial representation beyond the recent paper \cite{IKK15}.

In sect. 4 the quadratic truncation constraints are extracted from
(\ref{rll}) as eight relations in the tensor product formulation. The
decomposition in the spectral parameters leads to the natural ordering
of increasing complexity.
In the first step of transformations the constraints are separated into parts
symmetric or anti-symmetric in the tensor factors for further detailed
analysis. Further, the involved algebra valued matrices $G$ and $H$
are decomposed into parts graded anti-symmetric and symmetric in
matrix indices.
The anti-symmetric constraints imply commutation relations, in
particular the Lie algebra relations. All symmetric constraints can be written
in a standard form  and they imply relations  on the graded symmetric part
of the matrix expressions. Relations for the graded anti-symmetric parts are
derived from the anti-symmetric constraints.  Combining both  we
obtain the structure of the quadratic $L$ operator and
the final form of the  $\mathcal{Y}^{(2)}(\mathcal{G})$ algebra relations.

In sect. 5 we consider the reduced case where the second algebra valued
matrix $H$ is expressed completely in terms of the first $G$, i.e. all Yangian
generators are obtained from the ones  obeying the Lie algebra relations.

\section{Yangians of the orthogonal and  symplectic types}
\setcounter{equation}{0}

The fundamental Yang-Baxter equation has been written above, (\ref{ybe}).
It has a solution symmetric with respect to $so(n)$  or $sp(n)$
  \cite{ZZ,BKWK,Re,API}
    \be\label{rzW}
R_{b_1b_2}^{a_1a_2}(u)=u(u + \b) I^{a_1a_2}_{b_1b_2}+(u + \b)P^{a_1a_2}
_{b_1b_2}-\epsilon \, u \, K^{a_1a_2}_{b_1b_2} \,,
    \ee
    where
    \be
    \lb{IKP}
I^{a_1a_2}_{b_1b_2} = \delta^{a_1}_{b_1} \delta^{a_2}_{b_2}   \,, \qquad
P^{a_1a_2}_{b_1b_2} = \delta^{a_1}_{b_2} \delta^{a_2}_{b_1}   \,, \qquad
K^{a_1a_2}_{b_1b_2} = \varepsilon^{a_1a_2} \, \varepsilon_{b_1b_2} , \qquad
\b= \frac{n}{2} -\epsilon.
    \ee
Here $\e_{ab}$ is a non-degenerate invariant tensor, defining the scalar
product in $V$,
  \be
  \lb{meSOSp}
\e_{ab} = \epsilon \, \e_{ba}\, ,
\qquad\qquad  \e_{ab} \e^{bd} = \delta^d_a \; .
 \ee
The sign factor $\epsilon$ allows a uniform treatment of both orthogonal
and symplectic cases, $\epsilon= +1$ in the orthogonal and $\epsilon= -1$ in
the symplectic case. We shall call an expression graded symmetric if it is
symmetric in the orthogonal ($\epsilon = +1$) case and anti-symmetric in the
symplectic ($\epsilon = -1$) case.

The existence of the invariant tensor
$\varepsilon_{ab}$ leads to  the third term in the
corresponding expression of the $R$-matrices and  to the quadratic dependence
on the  spectral parameter $u$.

We consider now the $RLL$ relation (\ref{rll}) with
 $R^{a_1a_2}_{b_1b_2}(u-v)$  substituted by the Yang-Baxter
 $R$-matrix (\ref{rzW}). Further, substituting (\ref{defYan}) results in
the relations between the extended Yangian generators  $ (L^{(k)})^a_b $.
We shall investigate these relations in the case of truncation.

The fundamental Yang-Baxter equation has been written above, (\ref{ybe}),
in the index form referring to a basis in the tensor product of three copies of
the  $n$ dimensional linear space $V$, $V_1\otimes V_2\otimes V_3$. The
labels $1,2,3$ at the indices refer to the action on the corresponding tensor
factor. Alternatively, (\ref{ybe}) can be written in the standard auxiliary
space notation \cite{FRT} in terms of operators
acting in this tensor product space,
\be \label{ybe123}
R_{12}(u)R_{13}(u+v)R_{23}(v)=
 R_{23}(v)R_{13}(u+v)R_{12}(u).
\ee
where e.g.
 $R_{12}$ acts trivially on the third tensor factor and its matrix elements
with respect to a basis in $V_1\otimes V_2$
are denoted by  $ R_{b_1b_2}^{a_1a_2}$ as in (\ref{ybe}).
The expression (\ref{rzW}) for the $R$ matrix is
written in this notation in terms of $P_{12}, K_{12}$, the matrix elements
in a basis of $V_1\otimes V_2$ of which are as in (\ref{IKP}). They obey the
important relations
\be \lb{KP}
P_{12}K_{12}=\epsilon K_{12}=K_{12}P_{12}, \qquad \qquad K_{12}^2=n
\epsilon K_{12}.
\ee
In analogy, in this notation the $RLL$ relation (\ref{rll})   is formulated in
the product composed of the algebras of $End\; V_1 \otimes End \;V_2$ and the
Yangian algebra $\mathcal{Y}({\cal G})$,
\be \label{rll12}
 R_{12}(u-v) \, L_1(u) \, L_{2}(v)=
    L_{2}(v) \, L_{1}(u) \, R_{12}(u-v).
\ee
Here  $R_{12}$ has the unit element in the
$\mathcal{Y}({\cal G})$ factor,
$L_1$ has the unit element in the factor $End\; V_2$ and $L_2$
has the unit element in the factor $End\; V_1$.
The index form (\ref{rll}) is reconstructed from (\ref{rll12}) with
 the matrix  elements of $L_1, L_2$ with respect to a basis in $V_1 \otimes V_2$
written as
\be \label{L1L2}
 (L_1)^{a_1,a_2}{}_{b_1,b_2} = L^{a_1}_{\ b_1} \delta^{a_2}_{b_2}, \ \
(L_2)^{a_1,a_2}{}_{b_1,b_2}  =  \delta^{a_1}_{b_1} L^{a_2}_{\ b_2}.
\ee
The auxiliary space notation allows not only to write the basic relations in
a concise form but also to do effectively the transformations of the
truncated Yangian algebra relations. In the following we apply this
notation to transform the algebra relations into a form convenient to
understand how they constrain the representations.

The constraints to be obtained from the RLL relation (\ref{rll})
will be denoted by ${\mathfrak{C}}^{(i,k)}$ with the superscripts
$i=1, k=1,2,3$ in the case of linear evaluation and $i=2, k=1,...8$
in the case of quadratic evaluation. $C_1^{(i,k)}, C_2^{(i,k)}$ 
or $\bar C_1^{(i,k)}, \bar C_2^{(i,k)}$
will be used to abbreviate expressions related to the corresponding 
constraints ${\mathfrak{C}}^{(i,k)}$.
In the auxiliary space notation these expressions are algebra valued
matrices in $End\; V_1 \otimes End\; V_2$. As in the case of  $L_1, L_2$
the subscripts $1$ or $2$ mean that the expression has the unity in
$End\; V_2$ or $ End\; V_1 $ correspondingly.  $C^{(i,k)}$ (no subscript
$1,2$) is to be
understood as an algebra valued element of  $ End\; V $. We shall decompose
it with respect to index symmetry and have in particular trace contributions
denoted by $c^{(i,k)} I $, where $I$ is the unity in  $ End\; V $.

\section{Linear evaluation}
\setcounter{equation}{0}

We put all generators $(L^{(k)})^a_{\ b} \in \mathcal{Y}({\cal G})$ with $k>1$ equal to
zero and substitute in (\ref{rll}) the $L$-operator linear in the  spectral parameter:
 \be \label{L1}
  L^{a}{}_{b}(u) = u \delta^{a}_{b} + G^{a}{}_{b}.
    \ee
We formulate the known result which appeared already in \cite {SW,Re} and has
been considered in \cite{AMR,CDI1,IKK15}.

\begin{proposition}

 The Yangian linear evaluation $\mathcal{Y}^{(1)}(\mathcal{G})$
for $\mathcal{G} = so(n)$ and $\mathcal{G}= sp(n)$
is defined by the following algebra relations in terms of the algebra valued
matrix $G$ appearing in (\ref{L1}).

1. There is no traceless graded symmetric contribution in the matrix $G$, i.e.
 $$ G = g I + \bar G, \qquad tr\, G = n \cdot g, \qquad \bar G_{ba} = - \e
\bar G_{ab}. $$
$g$ is central and $\bar G$ is the graded antisymmetric part of the matrix
$G$.

2. The matrix elements of $\bar G$ obey the Lie algebra relations
of $so(n)$ ($ \e= +1$) or $sp(n)$ ($ \e= -1$), which can be written in
the auxiliary space notation as
$$ [\bar G_1 + P_{12}-\e K_{12}, \bar G_2] = 0. $$

3. The algebra relations are more restrictive compared to the Lie algebra
relations by the additional condition on the matrix  $\bar G$,
$$ \bar G^2 + \beta \bar G = m_2 \ I, \qquad m_2 = \frac{1}{n} tr \bar G^2.  $$

\end{proposition}


The proof extends over the next two subsections. The purpose is to illustrate
the methods in the simpler case of the linear evaluation
as a preparation to the application in the case of the quadratic evaluation.

\subsection{The conditions of truncation}

Here we start with the proof of Proposition 1 by obtaining the condition on
$G$ for the RLL relation (\ref{rll}) to hold.
With $L(u)$ substituted in the form (\ref{L1})
 the defining RLL-relation (\ref{rll}) takes the form:
$$
\Big(u(u+\b)I_{12}+(u+\b)P_{12}-u\epsilon K_{12}\Big)(u+v+G_1)(v+G_2)=
$$
$$
=(v+G_2)(u+v+G_1)\Big(u(u+\b)I_{12}+(u+\b)P_{12}-u\epsilon K_{12}\Big).
$$

The index form of $I_{12}$, $P_{12}$, $K_{12}$ is written above in (\ref{IKP}). The
index form of $G_1$ and $G_2$ is according to (\ref{L1L2})
$$
(G_1)^{a_1,a_2}{}_{b_1,b_2} = G^{a_1}{}_{b_1} \delta^{a_2}_{b_2},\qquad
\qquad  (G_2)^{a_1,a_2}{}_{b_1,b_2}=\delta^{a_1}_{b_1} G^{a_2}{}_{b_2}.
$$
The defining relation  can be further rewritten as
$$
(u+\b)\Bigl([G_1,G_2]+(G_1-G_2)P_{12}-\epsilon[K_{12},G_2]\Bigr)-
\epsilon v[K_{1 2},G_1+G_2]-
$$
$$
-\epsilon K_{12}(G_1-\b)G_2+\epsilon G_2(G_1-\b)K_{12}=0,
$$
and has to hold identically in  $u$ and $v$, implying
three restrictions on the matrix of generators $G$,
 \be\lb{c101}
-v{\mathfrak{C}}^{(1,1)}=-\epsilon v[K_{12},G_1+G_2]=0,
 \ee

 \be\lb{c102}
(u+\b){\mathfrak{C}}^{(1,2)}=(u+\b)\Bigl([G_1,G_2]+(G_1-G_2)P_{12}-
\epsilon[K_{12},G_2]\Bigr)=0,
 \ee
 \be\lb{c103}
-{\mathfrak{C}}^{(1,3)}=-\epsilon\Bigl(K_{12}(G_1-\b)G_2-G_2(G_1-\b)
K_{12}\Bigr)=0.
 \ee

We decompose the constraints with respect to the $1\leftrightarrow 2$
permutation parity.
$$
\mathfrak{C} = \mathfrak{C}_s+\mathfrak{C}_a, \qquad\qquad
P_{12} \mathfrak{C}_s P_{12} = \mathfrak{C}_s,\quad\quad
P_{12} \mathfrak{C}_a P_{12} = -\mathfrak{C}_a .
$$
Note that the first constraint is symmetric
$\mathfrak{C}^{(1,1)}=\mathfrak{C}^{(1,1)}_s$, while the second
is antisymmetric $\mathfrak{C}^{(1,2)}=\mathfrak
{C}^{(1,2)}_a$.
The antisymmetric part of the third constraint is
obtained by right and left multiplication by the projector
$\frac12(1-\epsilon P_{12})$.
 \be\lb{ac103}
{\mathfrak{C}}_{a,L}^{(1,3)}=K_{12}\Bigl([G_1,G_2]+\b(G_1-G_2)\Bigr)\!=
0=\!\Bigl([G_1,G_2]-\b(G_1-G_2)\Bigr)K_{12}={\mathfrak{C}}_{a,R}^{(1,3)}.
 \ee
However these relations do not imply a new constraint because they
follow from the second constraint upon multiplication by $K_{12}$.

The substantial part of the third relation, containing the new
restriction is obtained by symmetrization:
 \be\lb{sc103}
{\mathfrak{C}}^{(1,3)}_s=[K_{12},G_1^2+G_2^2]=0.
 \ee

\subsection{Analysis of the constraints}

We continue the proof of Proposition 1.

Multiplying the first constraint ${\mathfrak{C}}^{(1,1)}$
by $n\epsilon -K_{12}$ from the left and right, one obtains:
$$
\epsilon K_{12}(G_1+G_2)K_{12}=nK_{12}(G_1+G_2)=n
(G_1+G_2)K_{12}=2K_{12}{\rm{tr}}G,
$$
because by (\ref{KP}) $K_{12}G_1K_{12}=K_{12}G_2K_{12}=\epsilon K_{12}\ {\rm{tr}}G$.

In any case  $G$ can be separated into a traceless
part $\bar G$ and the trace part denoted by $g$:
 \be\lb{c111}
G=g I +\bar G.
\ee
Then the traceless part obeys
$$
\qquad K_{12}(\bar G_1+\bar G_2)=0=(\bar G_1+\bar
G_2)K_{12},\quad\Rightarrow\quad{\rm{tr}}\bar G\cdot K_{12}=0.
$$
The latter relation means
that the traceless graded-symmetric part of the
algebra-valued matrix $G$ vanishes. This can be checked by rewriting it in
the index notation.
Thus the first constraint  (\ref{c101}) implies that $G$
 consists of an arbitrary traceless,
graded-antisymmetric part $\bar G$ and the trace part, proportional to the metric tensor.
In the index notation this means:
$$ (g I)^a{}_b = g \delta^a_b, \ \  (g I)_{ab} = g \e_{ab}, $$
\be \label{antisym}
\bar G^a{}_a=0=\bar G_a{}^a,\qquad\qquad \bar
G_{c_1c_2}+\epsilon\bar G_{c_2c_1}=0=\bar G^{a_1a_2}+\epsilon\bar
G^{a_2a_1}.
\ee
Here repeated indices abbreviate the sum with the index running
over the index range and
$  \bar G^{a_1a_2} = \e_{a_1 a} \bar G^a{}_{a_2} $.

Substituting the solution (\ref{c111})  into
the  relations one concludes that the first constraint
${\mathfrak{C}}^{(1,1)}$ in terms of the traceless part $\bar G$ is
reduced to two:
  \be\lb{c121}
{\mathfrak{C}}^{(1,1)}_L=K_{12}(\bar G_1+\bar G_2)=0,\qquad
{\mathfrak{C}}^{(1,1)}_R=(\bar G_1+\bar G_2)K_{12}=0.
 \ee

The second constraint  is graded-antisymmetric (it has
a symmetry opposite to the one of the metric tensor).
 It expresses the symmetry of $G$ with respect to the simultaneous
rotation in the auxiliary   ({\it End} $V_1$ or {\it End} $V_2$)
and the quantum ($\mathcal{Y}^{(1)}$ representation) spaces and
can be rewritten as $\bar {\mathfrak{C}}^{(1,2)}$ in terms of $\bar
G$  as
  \be\lb{c152}
\bar{\mathfrak{C}}^{(1,2)}=[\bar G_1+P_{12}-\epsilon K_{12},\bar
G_2] =[\bar G_1,\bar G_2+P_{12}-\epsilon K_{12}].
 \ee
$\bar{\mathfrak{C}}^{(1,2)}=0$ is equivalent to the Lie algebra
commutation relations conventionally written as
\be \label{Lie}
[G_{ab}, G_{cd}]=-\e_{cb}G_{ad}+\e_{ad} G_{cb}+
\e_{ac} G_{bd} - \e_{db} G_{ca}.
\ee
(\ref{c152}) also shows that $g$ is central.

Multiplying $\bar {\mathfrak{C}}^{(1,2)}$  by $K_{12}$ from the left
and right and using the identities  $P_{12}K_{12}=\epsilon
K_{12}=K_{12}P_{12}$, $K_{12} ^2=n\epsilon K_{12}$ one obtains
a relation between the second and the third constraint,
 \be\lb{c122}
\bar {\mathfrak{C}}^{(1,2)}_L=K_{12}([\bar G_1,\bar G_2]-2\b\bar
G_2)= \bar {\mathfrak{C}}^{(1,3)}_{a,L},\qquad\bar
{\mathfrak{C}}^{(1,2)}_R=([\bar G_1,\bar G_2]+2\b\bar
G_2)K_{12}=\bar {\mathfrak{C}}^{(1,3)}_{a,L}.
 \ee
These relations  can be obtained by multiplication of (\ref{c103}) by
$1-\epsilon P_{12}$ from the left and from the right
correspondingly. Using (\ref{c121}) one can rewrite (\ref{c122}) as
 \be\lb{c132}
\bar {\mathfrak{C}}^{(1,2)}_L=K_{12}\Bigl((\bar G_1^2+\b \bar
G_1)-(\bar
 G_2^2+\b\bar G_2)\Bigr)=0,
 \ee
$$
\bar {\mathfrak{C}}^{(1,2)}_R=\Bigl((\bar
 G_1^2+\b \bar G_1)-(\bar G_2^2+\b\bar G_2)\Bigr)K_{12}=0.
 $$
These consequences of the second constraint tell that $\bar G^2 + \beta \bar G$
is graded-symmetric. This can be checked also in the index notation
on the basis of  (\ref{Lie}): From (\ref{antisym}) we see that $\bar
G^2_{ad} $ equals $ \e \bar G^2_{da} $ up to a commutator term, which can be
obtained from (\ref{Lie} ) by multiplying with $\e^{bc} $ and summing over
repeated indices. We obtain $ \bar G^2_{ad} = \epsilon \bar G^2_{da} - 2 \beta G_{ad}
$ and this implies the index symmetry relation 
$\bar G^2_{ad}+ \beta \bar G_{ad} = \epsilon (\bar G^2_{da}+ \beta \bar  G_{da} ) $.

The last symmetric constraint (\ref{sc103}) can be written as
 \be\lb{c153}
{\mathfrak{C}'}^{(1,3)}=[\epsilon K_{12},(\bar G_1^2+\b\bar
G_1)+(\bar G_2^2 +\b\bar G_2)]=0,
 \ee
and  allows us to deduce in analogy to the arguments leading to (\ref{c111})
and with the just established symmetry property of $\bar G^2 + \beta \bar G$
that  the latter term is just a trace contribution
 \be\lb{c163}
\bar G^2+\b\bar G= c^{(1.3)}I = m_2 \cdot I.
 \ee
Here $\bar G^2+\b\bar G$ is the graded-symmetric part of $G^2$  and
$m_2=\frac{1}{n} G^a{}_bG^b{}_a$ is the quadratic Casimir element.

The relation
(\ref{c163}) restricts the Lie algebra representation, which provides
the linear evaluation of the Yangian $\mathcal{Y}^{(1)}({\cal G})$.

This completes the proof of Proposition 1.

In this way, the Yangian linear evaluation $\mathcal{Y}^{(1)}(\mathcal{G})$
is defined by the following algebra relations: the
first constraint fixes the form of generator matrix (\ref{c111}), the next
one defines Lie algebra relations (\ref{c152}), while the last one
imposes the additional condition (\ref{c163}). Besides of the central
element $g$ the generators of $\mathcal{Y}^{(1)}(\mathcal{G})$ can be
considered as an image of the ones of the Lie
algebra. However the algebra relations for $\bar G$ are more restrictive by the
additional condition (\ref{c163}), which is not fulfilled identically in the
universal enveloping algebra $U(\mathcal{G})$.
 It can be fulfilled in distinguished $so(n)$ and $sp(n)$ representations only.
For example the generators
in the fundamental representation do not satisfy  this restriction.

Repeating the above analysis for the case of $g\ell(n)$ is much easier and we recover the
known statement that the linear evaluation of $\mathcal{Y}\big(g\ell(n)\big)$ always exists.
Indeed, in that case the fundamental $R$ matrix is simpler. It can be
obtained from (\ref{rzW}) by deleting the third term involving $K$. As a
consequence the additional constraint  (\ref{c163}) disappears.

\subsection{Spinorial Yang-Baxter  operators}

We consider the case that the generators are composed in terms
of an underlying algebra $\mathcal{C}$, which in turn is generated by
the elements $c^a$,
obeying the commutation relations of the oscillator algebra  or of the Clifford
algebra,
\be\lb{osc}
c^ac^b+\epsilon c^bc^a=\e^{ab},\qquad c_a=\e_{ab}c^b,\qquad
\Rightarrow\qquad c_ac^b+\epsilon c^bc_a=\d^b_a,
 \ee
$$
c_ac_b+\epsilon c_bc_a=[c_a,c_b]_{+\e}= \epsilon\e_{ab},\qquad\qquad c_ac^a=
\frac n2=\epsilon c^bc_b.
$$
We may consider $c^a$ as operators in the spinor space ($so(n)$ case) or
in  the Fock space of $\half n$ fermions ($so(n)$ case) or bosons ($sp(n)$ case).

Consider the linear map
$\rho : \mathcal{Y}^{(1)}(\mathcal{G}) \to \mathcal{C}$,
 \be\lb{Gosc}
\rho(G^a{}_b)=\tilde G^a{}_b =\frac\epsilon2\d^a_b-c^ac_b.
\ee
 We see that the image of $G$ is graded-antisymmetric,
i.e. $\tilde G = \rho (\bar G)$, $\rho(g)=0$, and check
that the Lie algebra  relations (\ref{c152})  are fulfilled.
Further,
$$ (\tilde G^2+\b \tilde G)^a{}_b=\frac\epsilon4(n-\epsilon)\d^a_b=
\frac\epsilon2(\b+\frac\epsilon2)\d^a_b.
 $$
Note that r.h.s. here fixes the value of the quadratic Casimir $m_2$ in this
particular representation.
We see that  the  composite generators (\ref{Gosc}) fulfill the additional condition
(\ref{c163}).
Thus the spinor representation (\ref{Gosc})
provides an example of the linear evaluation of the Yangian
$\mathcal{Y}^{(1)}({\cal G})$.


The Yang-Baxter operator $\check{\mathfrak{R}}$
intertwining two spinor representations, i.e. obeying
$$
\check{\mathfrak{R}}_{12}(u) \; {L} _1(u+v) \; {L} _2(v)=
 {L} _1(v) \; {L} _2(u+v) \; \check{\mathfrak{R}}_{12}(u) \; , \
$$
with
\be \label{Lcc}
 (L_1(u))^a_{\ b}=  \delta^a_b (u+\frac{\e}{2} ) - c_1^a c_{1 b} \ee
is known explicitly \cite{SW,CDI1,IKK15}.
In this Yang-Baxter relation the meaning of
 the indices $1,2$ is not literally the one of the auxiliary space notations used
everywhere else in the paper.
They refer here to two copies of the above
oscillator/ Clifford algebras, i.e. $V$ is replaced by the related Fock
space, and  $L_1 L_2$ means multiplication of fundamental representation
matrices.

Consider
the more general linear Yang-Baxter operator
\be \lb{spinL}
\mathcal{L}_1 (u) =
u \; I - \frac{1}{2} \, c_1^{[a} \, c_1^{b)} \otimes G_{ab}
\ee
where the fundamental representation has been replaced by some representation of the Lie
algebra with generators $ G_{ab}$. $[...)$ means the graded anti-symmetrization of
indices.
In \cite{CDI1} it has been established
that this form obeys the Yang-Baxter relation, obtained from the above by
replacing $L_1, L_2$ by $\mathcal{L}_1,  \mathcal{L}_2$,
\be \lb{spinrll}
\check{\mathfrak{R}}_{12}(u) \; \mathcal{L} _1(u+v) \; \mathcal{L} _2(v)=
 \mathcal{L} _1(v) \; \mathcal{L} _2(u+v) \; \check{\mathfrak{R}}_{12}(u) \; , \
\ee
 with the same spinorial $R$ operator if the additional condition
\be \lb{spincond1}
  \{ G_{a_1 [a_2} \, , \; G_{b_1 b_2)} \}  = 0 \ee
holds, where $[...)$ means the graded anti-symmetrization of indices
and $\{...,...\} $ means anticommutator.
Here the indices $1,2$ refer to two copies of the above oscillator/ Clifford
algebras as above, but  $\mathcal{L}_1 \mathcal{L}_2$
means multiplication in the algebra generated
by the matrix elements of $G$. Note that in (\ref{spinL}) only the graded
antisymmetric part of $G$ contributes. Therefore we identify $G$ with $\bar
G$ in the remaining part of this subsection.

\noindent

\begin{proposition}

 The additional condition (\ref{spincond1})
for a linear spinorial Yang-Baxter
operator $\mathcal{L}$ (\ref{spinL}) to obey (\ref{spinrll})
can be formulated in three other equivalent forms in terms of the
graded-antisymmetric matrix $G$ of generators obeying the $so(n)$ or $sp(n)$ Lie algebra
relations.
\begin{enumerate}

\item
\be \lb{spincond2}
  \{G_{a_1 a_2} \, , \; G_{b_1 b_2}\}  + \{G_{a_1 b_1} \, , \; G_{b_2 a_2}
\} + \{G_{a_1 b_2} \, , \; G_{a_2 b_1}\} = 0 . \ee

\item
\be \lb{spincond3}
W_{12} = 0, \ \ \
W_{12}=-( G_2+\epsilon)\Bigl((P_{12}-\epsilon K_{12}) G_2-
\epsilon G_1\Bigr).
 \ee
{\it $W_{12}$ is an algebra-valued element in $End V_1 \otimes End V_2$ with
matrix elements labeled by two index pairs,}
\be \label{W12}
(W_{12})_{a_1 b_1, a_2,b_2} = G_{[a_1 b_1} G_{a_2 b_2)}. \ee

\item
\be \lb{spincond4}
 {\hat G}^2 + \beta \hat G = c I , \ \ {\it where} \ \
 \hat G = -  \half c^{[a} \, c^{b)} \otimes G_{ab}, \ \ c= \frac{1}{8} \e
m_2 n . \ee
\end{enumerate}

 Further, the condition implies that the following cubic polynomial
in the matrix $G$ vanishes.
\be \lb{chi}
\chi(G)  = 0, \ \
\chi (z) =
z^3+(2\b+\epsilon)z^2+\epsilon(2\b -\frac{m_2}2) z-\frac{m_2}2 .
\ee
\end{proposition}

\noindent
{\bf Proof.}

The second form (\ref{spincond2}) is proved by writing the first form
(\ref{spincond1}) explicitly.
The third form (\ref{spincond3}) is proved by writing the matrix elements
of $W_{12}$  using (\ref{IKP}).
The fourth form is obtained by the expansion of the product \cite{IKK15}
$$ c^{[a} \, c^{b)}c^{[d} \, c^{e)} =
 c^{[a}  c^{b} c^{d}  c^{e)} - \half ( c^{[a} \, c^{e)} \varepsilon^{db}
+ c^{[b} \, c^{d)} \varepsilon^{ea} -  c^{[b} \, c^{e)} \varepsilon^{ad}-
 c^{[a} \, c^{d)} \varepsilon^{be}) +
\frac{1}{4} (\varepsilon^{ae} \varepsilon^{bd} -
\varepsilon^{eb}\varepsilon^{ad} ). $$
The expression on l.h.s of (\ref{spincond2}) can be written as the
graded-antisymmetric in indices sum of $G_{a_1 b_1}  G_{a_2 b_2}$.
This means, in $\hat G^2$ the contribution of the contraction with
$c^{[a_1}  c^{a_2} c^{b_1}  c^{b_2)}$ vanishes.

The statement about the cubic polynomial can be obtained by contraction with
a third factor of $G$, e.g. starting with the form (\ref{spincond3}) in terms
of $W_{12}$,
\be\lb{kgw}
K_{12}G_2W_{12} =-2\epsilon K_{12}\Bigl( G_2^3+(2\b+\epsilon)
G_2^2+\epsilon(2\b -\frac{m_2}2)
G_2-\frac{m_2}2\Bigr)\equiv-2\epsilon K_{12}\chi_2.
 \ee
$\chi_2$ denotes the expression in the bracket being the cubic polynomial
(\ref{chi}) with the argument substituted by $ G_2$.
Similarly one calculates:
 \be\lb{wgk}
W_{12} G_2K_{12}=-2\epsilon\chi_1 K_{12}.
 \ee
\hfill \qed

The notion of $W_{12}$ will be useful in the analysis of the quadratic
evaluation below. For this we notice the following properties:
  $W_{12}$ is annihilated by  $K_{12}$,
 \be\lb{wk}
W_{12}K_{12}=0=K_{12}W_{12}.
 \ee
In the index notation it means that the contraction of any pair of indices of $W^{a_2a_1}
{}_{c_1c_2}$  vanishes. Similarly,
$W_{12}$ is annihilated by  $P_{12}+\epsilon I_{12}$,
 \be\lb{kpe}
W_{12}P_{12}=P_{12}W_{12}=-\epsilon W_{12},\qquad\Rightarrow\qquad W_{21}=P_{12}W
_{12}P_{12}=W_{12}.
 \ee

The fourth form (\ref{spincond4}) written in terms of the spinorial matrix
$\hat G$ is reminiscent (but having an essentially different meaning)
to the additional condition of
the Yangian linear evaluation (\ref{c163}) written in the matrix $G$ in {\it{End
}}$V$, acting on the fundamental representation space $V$. However, the
condition that the linear spinorial $\mathcal{L}$ (\ref{spinL})
obeys the spinorial Yang-Baxter relation (\ref{spinrll}) results in the condition
on the latter matrix $G$
expressed instead in terms of the cubic polynomial $\chi(G)$ (\ref{chi}).
We shall see below that  such $G$, obeying
(\ref{spincond1}-\ref{spincond4}),
are appearing in a particular case of the quadratic evaluation.

\section{Quadratic evaluation}
\setcounter{equation}{0}

\subsection{The conditions of truncation}

Now we are going to consider the case where  $(L^{(k)})^a_{\ b} \in \mathcal{Y}({\cal G})$
with $k>2$ are constrained to vanish, i.e. we start from the quadratic ansatz
(\ref{Lo2}), $ L(u)=u^2 I +uG+H, $
and consider the conditions arising from (\ref{rll}) on the algebra-valued
matrices $G$ and $H$.

Note that in the resulting expressions for the constraints commutators and
anti-commutators will appear not graded by the dependence on $\e$, therefore
the notations $[...,...]$ and $\{...,...\}$ (no subscript) will be used,
respectively.


\begin{proposition}
The RLL relations imply 8 constraints. Four of them are symmetric and four
are anti-symmetric with repect to the permutation of the first two tensor
factors.  The four symmetric are ${\mathfrak{C
}}^{(2,1)}_s$, ${\mathfrak{C }}_s^{(2,3)}$, ${\mathfrak{C }}_s^{(2,6)}$
and ${\mathfrak{C }}^{(2,8)}_s$  and
the four anti-symmetric  are ${\mathfrak{C
}}_a^{(2,2)}$, ${\mathfrak{C}}_a^{(2,4)}=P_{12}{\mathfrak{C}}_a^{(2,5)}P
_{12}$ and ${\mathfrak{C}}_a^{(2,7)}$, where
 \be\lb{cs21}
{\mathfrak{C}}_s^{(2,1)}=
{\mathfrak{C}}^{(2,1)}=[P_{12}-\epsilon K_{12},G_1+G_2]=0,
 \ee
 \be\lb{ca22}
{\mathfrak{C}}_a^{(2,2)}=[G_1,G_2]-\frac12[P_{12}-\epsilon K_{12},G_1-G_2]=0,
 \ee
 \be\lb{cs23}
{\mathfrak{C}}_s^{(2,3)}=[P_{12}-\epsilon K_{12},H_1+H_2-\frac12
(G_1^2+G_2^2)]=0,
 \ee
 \be\lb{ca24}
{\mathfrak{C}}_a^{(2,4)}=[G_1,H_2]-[G_2,H_1]-[P_{12}-\epsilon K_{12},H_1-H_2]=0,
 \ee
 \be\lb{cs26}
{\mathfrak{C}}_s^{(2,6)}=[P_{12}-\epsilon K_{12},\{H_1,G_2\}+\{G_1,H_2\}],
 \ee
  \be\lb{ca27}
{\mathfrak{C}}_a^{(2,7)}=[H_1,H_2]+\frac14[P_{12}-\epsilon K_{12},
\{G_1,H_2\}-\{G_2,H_1\}],
 \ee
 \be\lb{cs28}
{\mathfrak{C}}_s^{(2,8)}=[P_{12}-\epsilon
K_{12},\{H_1,H_2\}-\b\epsilon(H_1+H_2)].
 \ee
\end{proposition}


The proof extends over this and the next subsections.
In this subsection we start analyzing the RLL relation (\ref{rll})
with the substitution (\ref{Lo2}) by decomposition in the spectral
parameters.
We shall complete the proof in the next
subsection by separation of the parts symmetric and anti-symmetric
in the labels $1,2$.

The defining relation (\ref{rll}) has the form:
 \be\label{rll2}
\big[u(u+\b)I_{12}+(u+\b)P_{12}-\epsilon uK_{12}\big]\big((u+v)^2+(u+v)G_1+
H_1\big)\big(v^2+vG_2+H_2\big)-
 \ee
$$
\!\!\!\!\!\!\!\!\big(v^2-vG_2+H_2\big)\big((u+v)^2-(u+v)G_1+H_1\big)\big
[u(u+\b)I_{12}+(u+\b)P_{12}-\epsilon uK_{12}\big]=0.
$$

This relation must hold at arbitrary values of the spectral
parameters $u$ and $v$, i.e. the coefficients at independent
monomials $u^kv^r$ must vanish. The l.h.s. of
(\ref{rll2})
 can be represented as a sum of the following eight
expressions.
 \be\lb{c201}
uv^2(u+v){\mathfrak{C}}^{(2,1)}=\epsilon uv^2(u+v)[K_{12},G_1+G_2],
 \ee
 \be\lb{c202}
(u+\b)uv(u+v){\mathfrak{C}}^{(2,2)}=(u+\b)uv(u+v)\Bigl([G_1,G_2]+
(G_1-G_2)P_{12}-\epsilon[K_{12},G_2] \Bigr),
 \ee
$$
-uv(u+v){\mathfrak{C}}^{(2,3)}=
$$
 \be\lb{c203}
=-\epsilon uv(u+v)\Bigl(K_{12} (H_1+H_2+(G_1-\b)G_2)
-(H_1+H_2+G_2(G_1-\b))K_{12}\Bigr),
 \ee
 \be\lb{c204}
-(u+\b)u(u+v){\mathfrak{C}}^{(2,4)}=-(u+\b)u(u+v)\Bigl([G_1,H_2]+
(H_1-H_2)P_{12}-\epsilon[K_{12},H_2] \Bigr),
 \ee
 \be\lb{c205}
-(u+\b)v(u+v){\mathfrak{C}}^{(2,5)}=-(u+\b)v(u+v)\Bigl([H_1,G_2]+
(H_1-H_2)P_{12}+\epsilon[K_{12},H_1] \Bigr),
 \ee
 \be\lb{c206}
uv{\mathfrak{C}}^{(2,6)}=\epsilon uv\Bigl(K_{12}(H_1(G_2+\b)+(G_1-
\b)H_2) -(H_2(G_1-\b)+(G_2+\b)H_1)K_{12}\Bigr),
 \ee
$$
-u(u+\b){\mathfrak{C}}^{(2,7)}=
$$
 \be\lb{c207}
=-u(u+\b)\Bigl([H_1,H_2]+(G_2H_1-H_2G_1) P_{12}-\epsilon
K_{12}(G_1-\b)H_2+\epsilon H_2(G_1-\b)K_{12}\Bigr),
 \ee
 \be\lb{c208}
-u{\mathfrak{C}}^{(2,8)}=-\epsilon u\Bigl(K_{12}(H_1-\b G_1+
\b^2)H_2 -H_2(H_1-\b G_1+\b^2)K_{12}\Bigr).
 \ee
They  are obtained as the result of the following arguments:
only two terms of (\ref{rll2}) are proportional to $uv^3$. After
extraction of (\ref{c201}), the only terms  proportional to
$u^3v$ are given by (\ref{c202}). Further, after extraction of
(\ref{c201}) and (\ref{c202}), the expression (\ref{rll2}) is at most
cubic in the  spectral parameters. Moreover, it contains $u^3$ only in
the combination (\ref{c204}) and $uv^2$  only in the combination
(\ref{c205}).
Extracting these combinations one can write
(\ref{rll2}) in the form quadratic in the spectral parameter $u$, which is
given by the sum of (\ref{c206}), (\ref{c207}) and (\ref{c208})

In other words, the spectral parameter dependence of the fundamental $R$
matrix and the quadratic $L$-operators obeying (\ref{rll})
can be translated into these
eight algebraic restrictions on the generators included in the
algebra-valued matrices $G$, $H$.

The first two constraints coincide with the ones,
appearing in the linear evaluation case, sect. 3.
The additional generators $H$ enter
now  the third relation $\mathfrak{C}^{(2,3)}$
and modify the symmetric part of $\mathfrak{C}^{(1,3)}$
lifting the restriction (\ref{c163}).
We shall see that the quadratic expression on l.h.s. of
(\ref{c163})
determines  the graded-symmetric part of  $H$.

\subsection{Permutation and index symmetry}

The permutation operator $P_{12}$ interchanging the order of the tensor
factors plays a crucial role.
As in the linear evaluation case we separate the symmetric and antisymmetric parts
of the constraints:
 \be\lb{syma}
{\mathfrak{C }}^{(2,k)}={\mathfrak{C }}_s^{(2,k)}+{\mathfrak{C
 }}_a^{(2,k)},\qquad k=1,\ldots 8,
 \ee
where ${\mathfrak{C}}_s^{(2,k)}=\frac12({\mathfrak{C}}^{(2,k)}+P_{12}
{\mathfrak{C}}^{(2,k)}P_{12})$ and ${\mathfrak{C}}_a^{(2,k)}=\frac12(
{\mathfrak{C}}^{(2,k)}-P_{12}{\mathfrak{C}}^{(2,k)}P_{12})$.
As the result of the truncation at second order we obtain
$2p=4$ symmetric constraints and $p^2= 4$ antisymmetric ones; the
situation is similar for the general case
$\mathcal{Y}^{(p)}(\mathcal{G})$ with the truncation at the order $p$.

The set of defining equations (\ref{c201}-\ref{c208}) is
equivalent to the following set of equations with definite symmetry
with respect to $1\leftrightarrow2$:

 \be\lb{c21}
{\mathfrak{C}}^{(2,1)}_a=0,\qquad{\mathfrak{C}}_s^{(2,1)}=
{\mathfrak{C}}^{(2,1)}=[P_{12}-\epsilon K_{12},G_1+G_2]=0,
 \ee
 \be\lb{c22}
{\mathfrak{C}}_s^{(2,2)}={\mathfrak{C}}_s^{(2,1)},\qquad {\mathfrak
{C}}_a^{(2,2)}=[G_1,G_2]-\frac12[P_{12}-\epsilon K_{12},G_1-G_2]=0,
 \ee
 \be\lb{acr23}
{\mathfrak{C}}^{(2,3)}_{a,L}=(1-\epsilon P_{12}){\mathfrak{C}}^{(2,3)}
=\Bigl([G_1,G_2]-\b(G_1-G_2)\Bigr)K_{12}={\mathfrak{C}}_a^{(2,2)}K_{12},
 \ee
 \be\lb{acl23}
{\mathfrak{C}}^{(2,3)}_{a,R}={\mathfrak{C}}^{(2,3)}(1-\epsilon P_{12})
=K_{12}\Bigl([G_1,G_2]+\b(G_1-G_2)\Bigr)=K_{12}{\mathfrak{C}}_a^{(2,2)},
 \ee
 \be\lb{sc23}
{\mathfrak{C}}_s^{(2,3)}=[P_{12}-\epsilon K_{12},H_1+H_2-\frac12
(G_1^2+G_2^2)]+\{{\mathfrak{C}}^{(2,1)},G_1+G_2\}=0,
 \ee
 \be\lb{sc24}
{\mathfrak{C}}_s^{(2,4)}=[G_1,H_2-\frac12G_2^2]+[G_2,H_1-\frac12G_1^2
]-{\mathfrak{C}}_s^{(2,3)}-\frac12\{G_1-G_2,{\mathfrak{C}}_s^{(2,2)}\},
 \ee
 \be\lb{ac24}
{\mathfrak{C}}_a^{(2,4)}=[G_1,H_2]-[G_2,H_1]-[P_{12}-\epsilon K_{12},H_1-H_2]=0,
 \ee
 \be\lb{ac26}
{\mathfrak{C}}^{(2,6)}_L={\mathfrak{C}}^{(2,6)}(1-\epsilon P_{12})=K_{12}
{\mathfrak{C}}_a^{(2,4)},\qquad {\mathfrak{C}}^{(2,6)}_R=(1-\epsilon
P_{12}){\mathfrak{C}}^{(2,6)}={\mathfrak{C}}_a^{(2,4)}K_{12},
 \ee
 \be\lb{sc26}
{\mathfrak{S}}_s^{(2,6)}=\frac12(1+\epsilon P_{12}){\mathfrak{C}}^{(2
,6)}(1+\epsilon P_{12})=[P_{12}-\epsilon K_{12},\{H_1,G_2\}+\{G_1,H_
2\}],
 \ee
 \be\lb{sc27}
{\mathfrak{C}}_s^{(2,7)}={\mathfrak{C}}_s^{(2,4)}P_{12}-\frac\epsilon
2\{K_{12},{\mathfrak{C}}_s^{(2,4)}\}-\frac\epsilon2{\mathfrak{C}}_s^
{(2,6)},
 \ee
 \be\lb{ac27}
{\mathfrak{C}}_a^{(2,7)}=[H_1,H_2]+\frac14[P_{12}-\epsilon K_{12},
\{G_1,H_2\}-\{G_2,H_1\}]-\frac\epsilon4\{K_{12},{\mathfrak{C}}_a^
{(2,4)}\},
 \ee
 \be\lb{ac28}
{\mathfrak{C}}^{(2,8)}_{a,L}={\mathfrak{C}}^{(2,8)}(1-\epsilon P_{12})=
\epsilon K_{12}\bigl({\mathfrak{C}}_a^{(2,7)}+\frac
{\epsilon-n}4{\mathfrak{C}}_a^{(2,4)}\bigr),
 \ee
 \be\lb{sc28}
{\mathfrak{C}}_s^{(2,8)}=[K_{12},\{H_1,H_2\}-\b\epsilon(H_1+H_2)]-
\frac\b2\{K_{12},{\mathfrak{C}}_s^{(2,4)}\}+\frac{\b\epsilon}2
{\mathfrak{C}}_s^{(2,6)},
 \ee
The calculation is straightforward using the identities (\ref{KP}).

We observe that the constraints ${\mathfrak{C}}_s^{(2,k)} $
with the labels $k=1,3,6,8$ contain the
corresponding symmetric constraint as their main parts while their
anti-symmetric parts appear as a consequence of the
 constraints ${\mathfrak{C}}_s^{(2,l)}$ with
$l<k$. In the constraints with the labels $k=2,4,5,7$ the
anti-symmetric parts are the leading contributions and the symmetric
ones  appear as a consequence of the
 constraints ${\mathfrak{C}}_s^{(2,l)}$ with
$l<k$. The anti-symmetric constraints involve commutators.

In this way we come to the independent constraints formulated in
Proposition 3 and thus we have completed the proof.


In the remaining part of this subsection we analyze the simpler constraints
with $k= 1,2,3,4,5$ and prove the following proposition.

\begin{proposition}
The $L$ operator representing the quadratic evaluation ${\mathcal{Y}}^{(2)}
({\mathcal{G}})$,
$$ 
L(u) = I u^2 + u G + H, 
$$
has the following structure in the decomposition of the  algebra-valued matrices
into trace contributions (proportional to $I$), graded-antisymmetric parts ($\bar G,
\bar H$) and  graded-symmetric parts. The traceless graded-symmetric part
$G$ vanishes. The traceless graded-symmetric part of $H$ equals the
traceless graded-symmetric part of $\bar G^2$.
$$
G= g I + \bar G, \qquad\qquad
H = hI + \bar H + \half (\bar G^2 + \beta \bar G).
$$
The matrix elements of $\bar G$ obey the Lie algebra relation of $so(n)$ or $sp(n)$ and
the further Yangian algebra generators in $\bar H$ transform as  the adjoint
representation of the latter.

\end{proposition}

\noindent
{\bf Proof.}

It is convenient to
decompose  $G$ and $H$ with respect to the index symmetry.
In sect. 3 we have seen this decomposition to appear for $G$ in analyzing
the first constraint in the first order evaluation case. In the case of
second order evaluation the frist constraint implies like in the first order
evaluation case  that  there is no
traceless graded-symmetric part in $G$,
$$
G= gI + \bar G.
$$
As in sect. 3 the trace contribution $g$ is central and $\bar G$ is
graded-antisymmetric.
Then the second constraint implies
 \be\lb{bc22}
\bar{\mathfrak{C}}^{(2,2)}=[\bar G_1+P_{12}-\epsilon K_{12},\bar
G_2]=0.
 \ee
It encodes the Lie algebra relations.
Multiplying (\ref{bc22}) by $K_{12}$ from the left and right, one
deduces  useful relations expressing the symmetry of $\bar G^2+\b\bar G$:
 \be\lb{symg2g}
K_{12}(\bar G_1^2+\b\bar G_1)=K_{12}(\bar G_2^2+\b\bar G_2),\qquad
(\bar G_1^2+\b\bar G_1)K_{12}=(\bar G_2^2+\b\bar G_2)K_{12}=0.
 \ee
It tells that $\bar G^2+\b\bar G$ is graded-symmetric. In the index notation
this is obtained from the graded-antisymmetry of $\bar G$ and the Lie algebra
commutation relations, which are contained in (\ref{bc22}).

The symmetric constraints $k=1, 3$ have the  standard form
 \be\lb{bc2k}
{\mathfrak{C}}_s^{(2,k)}=[K_{12},C^{(2,k)}_1+C^{(2,k)}_2]=0,\quad
k=1,3.
 \ee
The other symmetric constraints ($k=6,8$, not relevant in this proof)
can be written in this form too as will be shown in the next subsection.

This implies (by the same argument as in sect.3) that the expressions
$C^{(2,k)}$ have no traceless graded-symmetric contributions, i.e. decompose
into a trace contribution denoted by $c^{(2,k)}$  and a graded
antisymmetric matrix $\bar C^{(2,k)} $,
 \be\lb{cc2k}
C^{(2,k)}=c^{(2,k)} I +\bar C^{(2,k)},\qquad K_{12}(\bar C^{(2,k)}_1+
\bar C^{(2,k)}_2)=0=(\bar C^{(2,k)}_1+\bar C^{(2,k)}_2)K_{12}.
 \ee
The second relation in (\ref{cc2k}) does not fix the graded antisymmetric part.
In the cases
of $k= 1$ or $k=3$  this part is $\bar G$ or $ \bar H$ containing
independent algebra generators.

The first constraint ($k=1$) is analyzed completely above and the third
($k=3$) implies by the latter argument
the decomposition
 \be\lb{c223}
H=h I +\frac12(\bar G^2+\b\bar G)+\bar H,\quad K_{12}(\bar H_1+
\bar H_2)=0=(\bar H_1+\bar H_2)K_{12}\;\Leftrightarrow
\;\bar H_{ab}=-\epsilon\bar H_{ba},
 \ee
$h$ is the trace contribution proportional to unity matrix, $h= c^{(2,3)}$.
The graded-symmetric part of $H$ is fixed to be half of the graded symmetric part of
$\bar G^2$.

In terms of $\bar G, \bar H$ the graded anti-symmetric constraints with
$k=4,5$ read
 \be\lb{bc24}
\bar{\mathfrak{C}}^{(2,4)}=[\bar G_1+P_{12}-\epsilon K_{12},\bar
H_2]=0,
 \ee

 \be\lb{bc25}
\bar{\mathfrak{C}}^{(2,5)}=-P_{12}\bar{\mathfrak{C}}^{(2,4)}P_{12}=
[\bar H_1,\bar G_2+P_{12}-\epsilon K_{12}]=0.
 \ee
These relations tell  that the Yangian generators $\bar H$ transform under
the adjoint representation of the Lie algebra.
\hfill \qed

\subsection{Symmetric constraints}

We turn to the analysis of the more involved constraints,
the symmetric ones $k=6,8$ in this subsection and the antisymmetric
$k=7$ in the the following two subsections. This results in the algebra
relations for products of $\bar G$ and $\bar H$ formulated in the following

\begin{proposition}

The quadratic evaluation conditions result in the structure
formulated in Proposition 4
and further constrain the products of the algebra-valued
matrices $\bar G, \bar H$ as
\be\lb{ck11p}
\{\bar{H},\bar G\}+2\beta\bar{H}-g(\bar G^2+\b\bar G)=c^{(2,6)} \ I,
 \ee

\be \lb{bc7p}
[\bar H_1,\bar H_2]+\frac18[W_{12},\bar
G_1-\bar G_2]+\frac18[P_{12}-\epsilon K_{12},\chi_1-\chi_2-4g
(\bar H_1-\bar H_2)]+
\ee
$$
+\frac18  \alpha [P_{12}-\epsilon K_{12},\bar G_1-\bar G_2]=0,
$$
$\alpha = 4h+\b^2+1-2\epsilon\b+\frac{m_2}2\epsilon $,
 \be\lb{ck21p}
\bar{H}^2=c^{(2,8)} \ I +\frac14\bar G^4-g\b\bar H+\b\bar G^3+
(\frac54\b^2+h)\bar G^2+(\frac{\b^3}2+2h\b)\bar G.
 \ee
\end{proposition}

The proof extends over this and the next two subsections. Here we analyze the
symmetric constraints $k=6,8$. In the next subsection the needed information
from the anti-symmetric constraints is derived resulting in particular in a
relation for the commutator of the generators in $\bar H$.
This commutation relation will be reformulated in terms of the
  graded anti-symmetrized product of the generators $\bar G$ in the 5th
subsection.

We start the proof by recalling that the symmetric constraints ($k=1,3,6,8$)
can be written in the standard form
(\ref{bc2k}), which is to be show below for $k=6, 8$.
This implies (by the same argument as in sect.3) that the expressions
$C^{(2,k)}$ have no traceless graded-symmetric contributions, i.e. decompose
into a trace contribution denoted by $c^{(2,k)}$  and a graded
antisymmetric matrix $\bar C^{(2,k)} $, (\ref{cc2k}).

The second relation in (\ref{cc2k}) does not fix the graded antisymmetric part
of $\bar C^{(2,k)}$.
In the case
of $k= 1$ or $k=3$  this part is $\bar G$ or $ \bar H$ containing
independent algebra generators, i.e. the  constraints
do not imply relations expressing them in terms of a smaller set of elements.
In the other cases $k=6,8$ it takes to
derive from the anti-symmetric constraints a relation of the form
 $$ K_{12}(\bar C^{(2,k)}_1-\bar C^{(2,k)}_2)= 0 =
(\bar C^{(2,k)}_1-\bar C^{(2,k)}_2) K_{12}
$$
to fix it. This will be done in the next subsection.

We rewrite now all relations in terms of $\bar G, \bar H$ and the trace
contributions $g, h$.
 In particular, the  sixth
constraint takes the form:
 \be\lb{bc26}
-2\epsilon\bar{\mathfrak{C}}^{(2,6)}=[K_{12},\{\bar H_1,\bar  G_1\}
+\{\bar H_2, \bar G_2\}-g(\bar G_1^2+\bar G_2^2)]=0.
 \ee
This relation has now the standard form (\ref{bc2k}) with
 \be\lb{cc26}
C^{(2.6)}_1=\{\bar H_1,\bar  G_1\}-g(\bar
G_1^2+\b\bar G_1)+2\b\bar H_1.
 \ee
We have added the graded anti-symmetric term $\b(2\bar H-(\b+g)\bar G)$.

Let us transform finally the 8th constraint (\ref{cs28}) into
the standard form (\ref{bc2k}).
$$
{\mathfrak{C}}_s^{(2,8)}=[K_{12},\{\bar H_1,\bar H_2\}]+\frac12
[K_{12},\{\bar H_1,\bar G^2_2+\b\bar G_2\}+\{\bar G_1^2+
\b\bar G_1,\bar H_2\}]+
$$
$$
+\frac14[K_{12},\{\bar G_1^2,\bar G^2_2\}]+\frac\b2[K_{12},
\{\bar G_1^2,\bar G_2\}+\{\bar G_2^2,\bar G_1\}]+\frac{\b^2}4
[K_{12},\{\bar G_1,\bar G_2\}]+(h-\frac{\b\epsilon}2)
[K_{12},\bar G_1^2+\bar G_2^2].
$$
 We transform the terms containing $\bar H$:
$$
[K_{12},\{\bar H_1,\bar H_2\}]=[K_{12},-\bar H_1^2-\bar H_2^2].
$$
We show that terms linear in $\bar H$ cancel. Indeed,
$$
K_{12}(\{\bar G_1^2,\bar H_2\}+\{\bar G_2^2,\bar H_1\})
=K_{12}\Bigl((\bar G_1^2-\bar G_2^2)(\bar H_2-\bar H_1)+[\bar G_1,
\{\bar G_1,\bar H_1\}]+ [\bar G_2,\{\bar G_2,\bar H_2\}]\Bigr)=
$$
$$
=K_{12}\Big(\b(\bar G_1-\bar G_2)(\bar H_1-\bar H_2)-2\b
([\bar G_1,\bar H_1]+ [\bar G_2,\bar H_2])\Bigr)
=2\b K_{12}\Big(\bar H_1\bar G_1+\bar H_2\bar G_2\Bigr),
$$
Adding the other term containing $\bar H$  one obtains
$$
-\b K_{12}([\bar G_1,\bar H_1]+[\bar G_2,\bar H_2])=
\b K_{12}([\bar G_1,\bar H_2]+[\bar G_2,\bar H_1])=0,
$$
and similarly for terms right-multiplied by  $K_{12}$.

The terms containing only $\bar G$'s are
$$
\frac14[K_{12},(\bar G_1^2\bar G_2^2+\bar G_2^2\bar G_1^2)]-
\frac14[K_{12},\bar G_1^4+\bar G_2^4]=-\frac14[K_{12} ,(\bar G_1
^2-\bar G_2^2)^2]=\frac\b2[K_{12},\bar G_1^3+\bar G_2^3],
$$
and
$$
\frac\b4[K_{12}\{\bar G_1,\bar G_2^2\}+\{\bar G_2,\bar G_1^2\}]+
\frac{\b^2}4[K_{12}\{\bar G_1,\bar G_2\}]=
\frac\b4K_{12}(\bar G_1^2-\bar G_2^2)(\bar G_2-\bar G_1)-
$$
$$
-\frac\b4(\bar G_2-\bar G_1)(\bar G_1^2-\bar G_2^2)K_{12}-\frac
{\b^2}4[K_{12},\bar G_1^2+\bar G_2^2]
=\frac{\b^2}4[K_{12},\bar G_1^2+\bar G_2^2].
$$

Adding all contributions one obtains the standard form (\ref{bc2k})
with the consequences
 \be\lb{c248}
K_{12}\Bigl(-\bar H_1^2-\bar H_2^2+\frac14(\bar G_1^4+\bar
G_2^4)+\frac\b2(\bar G_1^3+\bar G_2^3)+(h+\frac{\b^2}4-
\frac{\b\epsilon}2)(\bar G_1^2+\bar G_2^2)\Bigr)\equiv 
K_{12}(C_1 ^{(2,8)} + C_2 ^{(2,8)} ) =
 \ee
$$
=K_{12}\Bigl(-\bar H_1^2-\bar H_2^2+\frac14(\bar G_1^4+\bar
G_2^4)+\b(\bar G_1^3+\bar G_2^3)+(h+\frac{5\b^2}4
)(\bar G_1^2+\bar G_2^2)\Bigr).
$$
$C_1 ^{(2,8)} + C_2 ^{(2,8)}$ abbreviates the expression in the bracket multiplying 
$K_{12}$.
Here we took into account the identity
$$
[K_{12},\bar G_1^3+\bar G_2^3+(\b+\frac n2)(\bar G_1^2+\bar G_2^2)]=0.
$$

\subsection{Anti-symmetric constraints}

The graded anti-symmetric constraints $k=2,4,5$ have been analyzed above
for the purpose of the proof of Proposition 4.

We intend to fix the graded anti-symmetric part of $C^{(2,6)}$,
 (\ref{cc26}).
The constraints $k=2,4$ provide the needed information.
Using the projections of the antisymmetric combination
 \be\lb{cl24}
\bar{\mathfrak{C}}_{a,L}^{(2,4)}=K_{12}\Bigl((\{\bar{H}_1,\bar G_1\}+
2\beta\bar{H}_1)-(\{\bar{H}_2,\bar G_2\}+2\beta\bar{H}_2)\Bigr)=0,
 \ee
and
 \be\lb{cr24}
\bar{\mathfrak{C}}_{a,R}^{(2,4)}=\Bigl((\{\bar{H}_1,\bar G_1\}+2\beta
\bar{H}_1)-(\{\bar{H}_2,\bar G_2\}+2\beta\bar{H}_2)\Bigr)K_{12}=0,
 \ee
as well as
 \be\lb{cl22}
\bar{\mathfrak{C}}_{a,L}^{(2,2)}=K_{12}\Bigl((\bar{G}_1^2+\b\bar G_1)-
(\bar{G}_2^2+\b\bar G_2)\Bigr)=0,
 \ee
and
 \be \lb{cr22}
\bar{\mathfrak{C}}_{a,R}^{(2,2)}
=\Bigl((\bar{G}_1^2+\b\bar G_1)-(\bar{G}_2^2+\b\bar G_2)\Bigr)K_{12}=0,
 \ee
one can rewrite this as
 \be\lb{c26a}
K_{12}(C^{(2,6)}_1-C^{(2,6)}_2)=0=(C^{(2,6)}_1-C^{(2,6)}_2)K_{12},
 \ee
due to (\ref{c132}). Combining (\ref{bc2k}), (\ref{cc26}) and (\ref{c26a}),
one deduces $[K_{12},C^{(2,6)}_1]=0=[K_{12},C^{(2,6)}_2]$. This means that
$C^{(2,6)}$, (\ref{cc26}),  has a trace contribution  proportional to $I$ only,
denoted by $c^{(2,6)}$,
 \be\lb{ck1}
\{\bar{H},\bar G\}+2\beta\bar{H}-g(\bar G^2+\b\bar G)=c^{(2,6)} I.
 \ee
Using this relation one can simplify the constraints (\ref{ca27}) and (\ref{cs28}).
We analyze the remaining graded antisymmetric  constraint $k=7$.
$$
{\mathfrak{C}}_a^{(2,7)}=[\bar H_1,\bar H_2]+\frac12([\bar H_1,\bar G_2^2]
+[\bar G_1^2,\bar H_2])+\frac\b2([\bar H_1,\bar G_2]+[\bar G_1,\bar H_2])
+\frac\b4([\bar G_1,\bar G_2^2]+[\bar G_1^2,\bar G_2])+
$$
$$
+\frac14[\bar G_1^2,\bar G_2^2]+\frac{\b^2}4[\bar G_1,\bar G_2]+
\frac14[P_{12}-\epsilon K_{12},(2h-g\b)(\bar G_1-\bar G_2)-2g(
\bar H_1-\bar H_2)-g(\bar G_1^2-\bar G_2^2)]+
$$
\be \label{C27a}
+\frac14[P_{12}-\epsilon K_{12},\{\bar G_1,\bar H_2\}-\{\bar G_2,\bar H_1
\}+\frac12(\{\bar G_1,\bar G^2_2\}-\{\bar G_2,\bar G_1^2\})].
\ee
First consider terms linear in $\bar H$.  Due to (\ref{bc24}-\ref{bc25}) we
have
$$
[\bar G_1,\bar H_2]=\frac12[P_{12}-\epsilon K_{12},\bar H_1-\bar H_2]=
[\bar H_1,\bar G_2],
$$
We rewrite also the last term  in the form
antisymmetric in $1\leftrightarrow2$  and obtain
$$
\frac12([\bar H_1,\bar G_2^2]+[\bar G_1^2,\bar H_2])+\frac14[P_{12}-
\epsilon K_{12},\{\bar G_1,\bar H_2\}-\{\bar G_2,\bar H_1\}]=\frac14
[P_{12}-\epsilon K_{12},\{\bar G_1,\bar H_1\}-\{\bar G_2,\bar H_2\}]=
$$
$$
=\frac14[P_{12}-\epsilon K_{12},-2\b(\bar H_1-\bar H_2)+g(\bar
G_1^2-\bar G_2^2)+g\b(\bar G_1-\bar G_2)].
$$
The contribution of the two other terms linear in $\bar H$ is
$$
\frac\b2([\bar H_1,\bar G_2]+[\bar G_1,\bar H_2])-\frac g2[P_{12}-
\epsilon K_{12},(\bar H_1-\bar H_2)]=\frac{\b-g}2[P_{12}-
\epsilon K_{12},(\bar H_1-\bar H_2)].
$$
Then we calculate the   contribution in (\ref{C27a}) cubic in $\bar G$,
$$
\frac14[\bar G_1^2,\bar G_2^2]+\frac18[P_{12}-\epsilon K_{12},
\{\bar G_1,\bar G^2_2\}-\{\bar G_2,\bar G_1^2\}]=\frac1{16}\{\bar
G_1^2+\bar G_2^2,[P_{12}-\epsilon K_{12},\bar G_1-\bar G_2]\}.
$$
The next term in  (\ref{C27a}) simplifies as
$$
\frac\b4([\bar G_1^2,\bar G_2]+[\bar G_1,\bar G_2^2])=\frac\b4
[P_{12}-\epsilon K_{12},\bar G_1^2-\bar G_2^2],
$$
and the remaining contributions in  (\ref{C27a}) are trivial.
Collecting all terms one obtains
$$
{\mathfrak{C}}_a^{(2,7)}=\bar{\mathfrak{C}}^{(2,7)}=[\bar H_1,\bar
H_2]+\frac1{16}\{\bar
G_1^2+\bar G_2^2,[P_{12}-\epsilon K_{12},\bar G_1-\bar G_2]\}+
$$
 \be\lb{bc27}
+\frac14[P_{12}-\epsilon K_{12},(2h+\frac{\b^2}2)(\bar G_1-\bar
G_2)-2g(\bar H_1- \bar H_2)+\b(\bar G_1^2-\bar G_2^2)].
 \ee
The constraint ${\mathfrak{C}}_a^{(2,7)}=0$ results in the condition
(\ref{bc7p}) of the proposition. In subsection 4.5 the expression will be
transformed in terms of $W_{12}$ (\ref{W12}).

In the last subsection we have derived from the 8th constraint a
relation for $\bar H^2$ leaving its graded anti-symmetric part undetermined.
To obtain this  part of $\bar H^2$ we multiply
 (\ref{bc27}) by $K_{12}$:
 \be\lb{kc7}
K_{12}{\mathfrak{C}}_a^{(2,7)}=K_{12}\Bigl(\bar H_1^2-\bar H_2^2+
\frac{\epsilon-\b}4(\bar G_1^3-\bar G_2^3)+\b g(\bar H_1- \bar H_2)-
 \ee
$$
-(\frac{\epsilon\b^2}4+
\frac{m_2}8+\b h)(\bar G_1- \bar G_2)\Bigr)=K_{12}(C_1^{(2,7)} -
C_2^{(2,7)}) =0.
$$
$C_1^{(2,7)} - C_2^{(2,7)}$ denotes the expression in the bracket multiplying $K_{12}$.
Similarly one obtains:
$$
{\mathfrak{C}}_a^{(2,7)}K_{12}=-K_{12}(C_1^{(2,7)} - C_2^{(2,7)})=0.
$$
Combining these relations and using the identities
 \be\lb{kg21}
K_{12}\bar G_2=-K_{12}\bar G_1,\qquad K_{12}\bar H_2=-K_{12}\bar H_
1,\qquad K_{12}(\bar G_2^2+\b\bar G_2)=K_{12}(\bar G_1^2+\b\bar G_1),
 \ee
 \be\lb{kg32}
K_{12}\Bigl(\bar G_2^3+(\b+\frac n2)\bar G_2^2-\frac{m_2}2\Bigr)=
-K_{12}\Bigl(\bar G_1^3+(\b+\frac n2)\bar G_1^2-\frac{m_2}2\Bigr),
 \ee
 \be\lb{kg431}
K_{12}\Bigl(\bar G_1^4-\bar G_2^4+(\epsilon+3\b)(\bar G_1^3-\bar G_2^3)
-(\b^2(\epsilon+3\b)+\frac{m_2}2)(\bar G_1-\bar G_2)\Bigr)=0,
 \ee
one can rewrite (\ref{kc7}) in a form convenient for comparison with (\ref{c248}) as
$$
K_{12}\Bigl(\bar H_1^2-\bar H_2^2-\frac14(\bar G_1^4-\bar G_2^4)+
g\b(\bar H_1-\bar H_2)-\b(\bar G_1^3-\bar G_2^3)-
$$
$$
-(\frac54\b^2+h)(\bar G_1-\bar G_2)-(\frac{\b^3}2+2h\b)(\bar G_1-
\bar G_2)\Bigr)= K_{12} (C_1^{(2,8)} - C_2^{(2,8)}) = 0.
$$
In this way we determine $\bar H^2$ up to a trace part involving  $c^{(2,8)}$.
 \be\lb{ck2}
\bar{H}^2=c^{(2,8)}I +\frac14\bar G^4-g\b\bar H+\b\bar G^3+
(\frac54\b^2+\frac{\b g}2+h)\bar G^2+(\frac{\b^3}2+\frac{\b^2g}2+2h\b)\bar G.
 \ee

\subsection{The seventh constraint in terms of $W_{12}$}

We recall $W_{12}$ which is the graded-antisymmetrization in indices
of $G_1G_2$ (\ref{spincond3}),
$$W_{12}=-(\bar G_2+\epsilon)\Bigl((P_{12}-\epsilon K_{12})\bar G_2-
\epsilon\bar G_1\Bigr),
$$
and its relations (\ref{kgw} - \ref{wk}).

We start with the second term in the  seventh constraint (\ref{bc27})
and use  the notation as in (\ref{kc7}) :
 \be\lb{con7}
C_7=\frac1{16}\{\bar G^2_1+\bar G^2_2,[P_{12}-\epsilon K_{12},\bar
G_1-\bar G_2]\}=
 \ee
$$
=\frac18\Big([P_{12}-\epsilon K_{12},\bar G_1^3]-\bar G_1[P_{12}-
\epsilon K_{12},\bar G_1]\bar G_1\Big)-1\leftrightarrow2.
$$
Consider the commutator
 \be\lb{wgw}
[\bar G_2+\epsilon,W_{12}]=(\bar G_2+\epsilon)[P_{12}-\epsilon
K_{12},\bar G_2](\bar G_2+\epsilon),
 \ee
 Multipling it by $P_{12}$ from both sides we have also
 \be\lb{wgwp}
[\bar G_1+\epsilon,W_{12}]=(\bar G_1+\epsilon)[P_{12}-\epsilon
K_{12},\bar G_1](\bar G_1+\epsilon),
 \ee
due to the symmetry   $W_{21}=W_{12}$ (\ref{kpe}). This implies
 \be\lb{wgw12}\!
[\bar G_1-\bar G_2,W_{12}]=\bar G_1[P_{12}-\epsilon
K_{12},\bar G_1]\bar G_1+\epsilon[P_{12}-\epsilon
K_{12},\bar G_1^2]+[P_{12}-\epsilon
K_{12},\bar G_1]-1\leftrightarrow2.
 \ee
We see that some terms of (\ref{con7}) appear here.
Multiplying this expression  by $K_{12}$
one obtains using $K_{12}W_{12}=0$
$$
K_{12}(\bar G_1-\bar G_2)W_{12}\!=\!K_{12}\!\Big(\!\epsilon(2(\bar G_2^3
-\bar G_1^3-\b(\bar G_1^2-\bar G_2^2))+[\bar G_1^2,\bar G_2]+
[\bar G_1,\bar G_2^2])+(m_2-2\b)(\bar G_1-\bar G_2)\!\Big)\!=
$$
$$
=K_{12}\Big(2\epsilon(\bar G_2^3-\bar G_1^3)+2\b(\epsilon+1)\bar G_1^2
-\bar G_2^2)+(m_2-4\b)\epsilon(\bar G_1-\bar G_2)\Big)=
$$
$$
=-2\epsilon K_{12}\Big(\bar G_1^3-\bar G_2^3+(2\b-\frac{m_2}2-\b(2\b+
\epsilon)(\bar G_1-\bar G_2)\Big)\equiv-2\epsilon K_{12}(\chi_1-\chi_2).
$$
Similarly, the multiplication by $K_{12}$ from the right leads to
$$
W_{12}(\bar G_1-\bar G_2)K_{12}=-2\epsilon\Big(\bar G_1^3-\bar G_2^
3+(2\b-\frac{m_2}2-\b(2\b+\epsilon)(\bar G_1-\bar G_2)\Big) K_{12}=-2
\epsilon (\chi_1-\chi_2)K_{12}.
$$
Here $\chi$ is given by (\ref{chi}).
Then
$$
C_7=\frac18[W_{12},\bar G_1-\bar G_2]+\frac18[P_{12}-\epsilon
K_{12},\chi_1-\chi_2 -2\b(\bar G_1^2-\bar G_2^2)+(1-
2\epsilon\b+\frac{m_2}2\epsilon)(\bar G_1-\bar G_2)],
$$
and hence according to (\ref{bc27})
$$
{\mathfrak{C}_a}^{(2,7)}=[\bar H_1,\bar H_2]+\frac18[W_{12},\bar
G_1-\bar G_2]+\frac18[P_{12}-\epsilon K_{12},\chi_1-\chi_2-4g
(\bar H_1-\bar H_2)]+
$$
$$
+ \alpha \frac18[P_{12}-\epsilon K_{12},\bar G_1-\bar G_2]=0.
$$
$\alpha = 4h+\b^2+1-2\epsilon\b+\frac{m_2}2\epsilon $.

This completes the proof of Proposition 5.


\subsection{The center of the truncated Yangian ${\cal{Y}}^{(p)}(\cal{G})$}

The center of the extended Yangian algebra of
the orthogonal and symplectic types
has been analyzed in  \cite{AMR} and formulated in terms of the reproducing
function. In order to define it
we consider again the RLL-relation (\ref{rll}) with the fundamental $R$-matrix
given by (\ref{rzW}) at $u=v-\b$. We  arrive at
\be\lb{kll}
K_{12}L_1(v-\b)L_2(v)=L_2(v)L_1(v-\b)K_{12}.
\ee
In the index notation this reads
\be\lb{kll1}
\e^{a_1a_2}\e_{b_1b_2}L^{b_1}{}_{c_1}(u-\b)L^{b_2}{}_{c_2}(u)=
L^{a_2}{}_{b_2}(u)L^{a_1}{}_{b_1}(u-\b)\e^{b_1b_2}\e_{c_1c_2}.
\ee
After multiplication by $\e_{a_2a_1}$ we obtain
\be\lb{kll2}
C_{ab}(u)\equiv L_{ca}(u-\b)L^c{}_b(u)=\frac1n\e_{ab}
L_c{}^d(u)L^c{}_d(u-\b)\equiv \e_{ab}c(u),
\ee
where according to \cite{AMR} the center reproducing function is defined as
\be\lb{c}
C(u)=L^t(u-\b)L(u)=c(u)I.
\ee
 One proves that $C(u)$ contains central elements by showing that it
commutes with $L(v)$,
$$
C(u)L_2(v)=L^t_1(u-\b)L_1(u)L_2(v)=L^t_1(u-\b)R^{-1}_{12}(u-v)L_2(v)
L_1(u)R_{12}(u-v)=
$$
$$
=L_2(v)R^{-1}_{12}(u-v)L^t_1(u-\b)L_1(u)R_{12}(u-v)=L_2(v)R^{-1}_{12}
(u-v)C(u)R_{12}(u-v)=L_2(v)C(u),
$$
here in the last step we have used (\ref{c}).

In the case of the linear ansatz $L(u)=uI+G$ the reproducing function looks like:
\be\lb{c1}
C^{(1)}_{ab}=\e_{ad}{C^{(1)}}^d{}_b(u)=\e_{ad}\e_{ce}\big((u-\b)\d^e_f+
G^e{}_f\big)\e^{df}(u\d^c_b+G^c{}_b)=
\ee
$$
=\e_{ce}\big((u+g-\b)\d^e_a+\bar G^e{}_a\big)((u+g)\d^c_b+\bar G^c{}_b)=
$$
$$
=\epsilon\Big((u+g)(u+g-\b)\e_{ab}+(u+g-\b)\bar G_{ab}-(u+g)\bar G_{ab}-
\bar G_{ac}\bar G^c{}_b\Big)=
$$
$$
=\epsilon\Big((u+g)(u+g-\b)\e_{ab}-(\bar G^2+\b\bar G)_{ab}\Big)=
\epsilon\big((u+g)(u+g-\b)-c^{(1.3)}\big)\e_{ab}.
$$
Here we have used the notation (\ref{c163}) for $c^{(1.3)} = g$ to emphasize
the analogy with the quadratic evaluation case below.
Note that all graded antisymmetric terms (like $\bar G$)  cancel out
and the condition that $C(u)$ is central reproduces the linear evaluation
constraint.

That $g$ commutes with $G$ is evident from the second constraint
in both the linear (\ref{c102}) and the quadratic (\ref{c22})
evaluation cases.
The center
reproducing function allows us to prove easily,
that  in the latter case
 $g$ commutes with $H$ and
$h,  c^{(2.6)}, c^{(2.8)}$  are central too.

For this we substitute  $L(u)$ by the quadratic ansatz (\ref{Lo2})
into (\ref{kll2}).
$$
C^{(2)}_{ab}(u)=\e_{ad}{C^{(2)}}^d{}_b(u)=\e_{ad}\e_{ce}\big((u-\b)^2\d
^e_f+(u-\b)G^e{}_f+H^e{}_f\big)\e^{df}(u^2\d^c_b+uG^c{}_b+H^c{}_b)=
$$
\be\lb{cc2}
=\epsilon\Big((u-\b)^2\e_{ac}+(u-\b)(g\e_{ac}-\bar G_{ac})+(h\e_{ac}
+\frac12(\bar G^2+\b\bar G)_{ac}-\bar H_{ac})\Big)\Big(u^2\d^c_b+
\ee
$$
+u(g\d^c_b+\bar G^c{}_b)+(h\d^c_b+\frac12(\bar G^2+\b\bar G)^c{}_b+
\bar H^c{}_b\Big).
$$
Here at the first step we  have used $\e_{ad}\e^{df}=\d^f_a$, lowered the index
$f$ by using the metric tensor. We have taken into account the symmetry properties:
$\e_{ca}=\epsilon\e_{ac}$, $\bar G_{ca}=-\epsilon\bar G_{ac}$, $\bar
H_{ca}=-\epsilon\bar H_{ac}$ and $(\bar G^2+\b\bar G)_{ca}=\epsilon
(\bar G^2+\b\bar G)_{ac}$.
Now we omit the matrix indices and collect similar terms:
$$
\epsilon C^{(2)}(u)=((u-\b)^2+(u-\b)g+h)(u^2+ug+h)+(g(u-\b)+\frac{
\b^2+g\b}2+h)(\bar G^2+\b\bar G)+
$$
$$
+\b h\bar G+(\b^2-2u\b-g\b)\bar H+\frac14(\bar G^4+4\b\bar G^3
+3\b^2\bar G^2)-u\bar H\bar G-(u-\b)\bar G\bar H+\frac12[\bar G^2
+\b\bar G,\bar H]-\bar H^2.
$$
We take into account that $[\bar G,\bar H]=\b\bar H$ and obtain
$$
\epsilon C^{(2)}(u)=(h+(u-\b)^2+(u-\b)g)(h+u^2+ug)+(\b-u)\Big(\{\bar
G,\bar H\}+2\b\bar H-g(\bar G^2+\b\bar G)\Big)+
$$
\be\lb{cc21}
+\Big(-\bar H^2-g\b\bar H+\frac14\bar G^4+\b\bar G^3+(\frac54\b^2+
\frac{g\b}2+h)\bar G^2+(\frac{\b^3}2+\frac{g\b^2}2+2h\b)\bar G\Big)=
\ee
$$
=(u^2+ug+h)(h+(u-\b)^2+(u-\b)g)+(\b-u)c^{(2.6)}-c^{(2.8)}.
$$
At the last step we have noticed that in the large-bracket terms
the expressions of the conditions (\ref{ck1}) and (\ref{ck2}) appear.


\section{ The Lie algebra resolution}
\setcounter{equation}{0}

We consider the case where the second non-trivial term in
the quadratic $L$ (\ref{Lo2}) is completely expressed in terms of the
 generators in $G$ obeying the Lie algebra relations. This restriction results in the
Lie algebra resolution of the Yangian second order evaluation.
Formally this can be understood as a map $\rho$ of
$ \mathcal{Y}^{(2)}(\mathcal{G}) $ to the associative algebra generated by
the matrix elements of $G$,
\be\label{hhh}
\rho(\bar H) =a\bar G,\qquad\qquad \rho(G) = G=g I +\bar G,
\ee
where $a$ is a central element.
 \be\label{ggg}
\rho(H) = h I +\frac12\bar G^2+(a+\frac\b2)\bar G.
 \ee
We shall see that in this case the algebra relations are fulfilled if
a simple condition on the product of the generators in $G$ holds. Then
$a$ and also $g$ and $h$ are fixed. This condition is related to
the vanishing of a third order polynomial in $\bar G$, in analogy to the linear
evaluation case, where a condition is the vanishing of a second order
polynomial in $\bar G$.  Higher order evaluations are related to
such polynomials of corresponding higher order.
In the first subsection we discuss, how polynomial conditions in $\bar G$
constrain the algebra, assuming that $\bar G$ obeys the Lie
algebra relations.

\subsection{Representations specified by characteristic
polynomials}

Let us consider representations of the orthogonal or symplectic Lie
algebra $\mathcal{G}$ constrained by the vanishing of
 a polynomial in $\bar G$.

 If this  constraint is  of first order in $\bar G$ it does not
allow non-trivial representations.
 Consider the quadratic case.
$$
\chi^{(2)}=\bar G^2+A\bar G+B.
$$
and discuss which values of the coefficients $A, B$ may be chosen.

 This expression can be
represented as a sum of graded symmetric and antisymmetric matrices,
$$
\chi^{(2)}_{ab}={\chi^{(2)}_s}_{ab}+{\chi^{(2)}_a}_{ab},\qquad\qquad
{\chi^{(2)}_s}_{ab}=\frac12(\chi^{(2)}_{ab}+\epsilon\chi^{(2)}_{ab}),
\qquad{\chi^{(2)}_a}_{ab}=\frac12(\chi^{(2)}_{ab}-\epsilon\chi^{(2)}_{ab}),
$$
and the condition $\chi^{(2)}= 0$ implies both parts to vanish,
${\chi^{(2)}_a}=0$ and ${\chi^{(2)}_a}=0$. Let us rewrite the two conditions
in terms of the tensor product notation,
$$
K_{12}(\chi^{(2)}_{s 1}+\chi^{(2)}_{s 2})=0,\qquad\qquad
 K_{12}(\chi^{(2)}_{a 1}-\chi^{(2)}_{a 2}) =0.
$$

Due to the Lie
algebra relations the antisymmetric part $\chi^{(2)}_a$  is given by the
polynomial of lower  order, because
$$K_{12}(\bar G^2_1-\bar G^2_2)=-\b K_{12}(\bar G_1-\bar G_2) . $$
In order to avoid a restriction of the first order in $\bar G$ we have to
specify the coefficients such that the graded anti-symmetric part vanishes.
$$
\chi^{(2)}_s=\bar G^2+\b\bar G -m_2I,
$$
where $m_2$ stands for the quadratic Casimir and $I$ is the unit matrix.

Higher order polynomial constraints are to be analyzed  analogously.
One finds that the order $p$ constraint should be
graded symmetric if $p$ is even and graded anti-symmetric if $p$ is odd.
Otherwise a constraint of order $p-1$ would be involved.

Let us address the case of current interest and
consider an arbitrary cubic polynomial in $\bar G$
$$
\chi^{(3)}=\bar G^3+D\bar G^2+E\bar G+F.
$$
We decompose again into graded  symmetric and antisymmetric
parts,
$$
K_{12}\chi^{(3)}_1=\frac12K_{12}(\chi^{(3)}_1-\chi^{(3)}_2)+
\frac12K_{12}(\chi^{(3)}_1+\chi^{(3)}_2).
$$
The symmetric part is reduced to the second order polynomial,
$$
\frac12K_{12}(\chi^{(3)}_1+\chi^{(3)}_2)=\frac12K_{12}\Big([\bar
G_1,\bar G_2](\bar G_1-\bar G_2)+\bar G_1[\bar G_1,\bar G_2]
+D(\bar G_1^2+\bar G_2^2)+2F\Big)=
$$
$$
=\frac12K_{12}\Big((D-2\b-\epsilon)(\bar G_1^2+\bar G_2^2)+2F+m_2\Big).
$$
In turn the antisymmetric part contains only one free parameter,
$$
\frac12K_{12}(\chi^{(3)}_1-\chi^{(3)}_2)=\frac12K_{12}\Big(\bar
G_1^3-\bar G_2^3+(E-D\b)(\bar G_1-\bar G_2)\Big).
$$
We should not allow $\chi^{(3)}$ to include a graded symmetric part,
this  means $2F=-m_2$ and $D=\epsilon+2\b$.
 In this way one deduces
that the cubic polynomial appropriate for a constraint
has the form
 \be\label{cpo}
\chi^{(3)}=\bar G^3+(\epsilon+2\b)\bar G^2+E\bar G-\frac{m_2}2.
 \ee
with one free parameter $E$. It means that the graded anti-symmetric part of
the third power $\bar G^3$ is constrained to be proportional to the first
power $\bar G$.

Now the comparison with (\ref{chi}) shows that the free parameter $E$ in
our polynomial $\chi$ related to $W_{12}$ is given by
$$
E=\b D+2\b^2+\epsilon\b-\frac{m_2}2=\epsilon(2\b-\frac{m_2}2),
$$
and
 \be\label{chi4}
\chi=\bar G^3+(\epsilon+2\b)\bar G^2+(2\epsilon\b-\frac{
\epsilon m_2}2)\bar G-
\frac{m_2}2.
 \ee

 \subsection{The $W_{12}$ condition}

\begin{proposition}

 The conditions of the Lie algebra resolution of the second order
Yangian evaluation, where $L(u)$ has the form
\be \label{L2G}
L(u)=u^2+u(g+\bar G)+h+\frac12(\bar G^2+(\b+g)\bar G),
\ee
are fulfilled if the matrix $\bar G$ obeying the Lie algebra
relation obeys additionally the condition of vanishing of its graded
anti-symmetrized product,
\be \label{W125}
 (W_{12})_{a_1 b_1a_2b_2} = G_{[a_1 b_1} G_{a_2 b_2)} = 0, \ee
and the central elements are all expressed in terms of $ m_2
= \frac{1}{n} tr(\bar G^2)$ as
 \be\lb{gh7}
g^2=-\b^2-\frac{m_2}8,\qquad 4h=2\b^2-1+2\b\epsilon-\frac{m_2}8.
 \ee
\end{proposition}

\noindent
{\bf Proof.}

 With the   above restrictions (\ref{hhh}) the first five
constraints (\ref{cs21}-\ref{ca24}) hold  and we have to
 check the remaining three. After the substitution of (\ref{hhh})
and (\ref{ggg}) into (\ref{cs26}) the latter takes the form
 \be\label{k26}
[K_{12},\frac12\{\bar G_1^2,\bar G_2\}+\frac12\{\bar G_2^2,\bar G_1\}
+g(\bar G_1^2+\bar G_2)+(2a+\b)\{\bar G_1,\bar G_2\}]=0.
 \ee
We consider the first term in the commutator.
$$
\frac12K_{12}\Big(\{\bar G_1^2,\bar G_2\}+\{\bar G_2^2,\bar G_1
\}\Big)=K_{12}[\bar G_1,\bar G_2](\bar G_2-\bar G_1)=\frac\b2K_
{12}(\bar G_1-\bar G_2)^2=\b K_{12}(\bar G_1^2+\bar G_2^2),
$$
$$
\frac12\Big(\{\bar G_1^2,\bar G_2\}+\{\bar G_2^2,\bar G_1\}\Big)
K_{12}=(\bar G_1-\bar G_2)[\bar G_1,\bar G_2]K_{12}=\frac\b2
(\bar G_1-\bar G_2)^2K_{12}=\b (\bar G_1^2+\bar G_2^2)K_{12}.
$$
Thus (\ref{k26}) acquires the form
 \be\label{k126}
(g-2a)[K_{12},\bar G_1^2+\bar G_2^2]=0.
 \ee

The relation (\ref{bc26}) then tells that
$$
2a=g, \ \ c^{(2,6)} = 0.
$$
Substituting $\bar H=\frac g2\bar G$ into (\ref{bc7p})
results in
 \be\label{k277}
{\mathfrak{C}_a}^{(2,7)}=\frac18[W_{12},\bar
G_1-\bar G_2]+\frac18[P_{12}-\epsilon K_{12},\chi_1-\chi_2]
+\frac{\a'}8[P_{12}-\epsilon K_{12},\bar G_1-\bar G_2],
 \ee
$\a'=\alpha+3g^2
= 4h+\b^2+1-2\epsilon\b+\frac{m_2}2\epsilon+3g^2 $.

Because $\chi(\bar G)$ is a contraction  of $W_{12}$ with $\bar G$
(see proof of proposition 1)
this constraint is fulfilled if
\be \lb{W0}
 W_{12}= 0 , \qquad\qquad\alpha\p = 0. \ee

Finally, we turn to the eighth constraint.
Substituting $\bar H = \half g \bar G$ in (\ref{ck2}) we obtain
the fourth order polynomial condition in the algebra valued matrix  $\bar G$.
\be \lb{r8}
\bar G^4 + 4\beta \bar G^3 + (5\beta^2 +4h-g^2) \bar G^2 + 2 \beta
(\beta^2 +4h-g^2) \bar G + 4 c^{(2,8)} = 0.
\ee
This is compatible with the sufficient condition for solving the
seventh reduced constraint (\ref{W0}) only if
$$
4h-g^2=3\b^2+2\b\epsilon-1,\qquad\qquad4c^{(2,8)}=\frac{m_2}2
(\epsilon-2\b).
$$
Indeed, with these relations between central elements
the 4th order polynomial on l.h.s of (\ref{r8}) is expressed in terms of
$\chi$ (\ref{chi4}) as
$$(\bar G+2\b-\epsilon) \chi(\bar G). $$
This means that the 8th constraint is fulfilled if (\ref{W125})
holds.
\hfill \qed

The condition of the vanishing $W_{12}$ appeared in subsect. 3.3
and has been shown to be  equivalent to the one
found in \cite{CDI1} for the
linear spinorial Yang-Baxter operator $\mathcal{L}$ (\ref{spinL})
to obey the spinorial RLL relation (\ref{spinrll}).

We remind the Jordan-Schwinger example of second order evaluation
which first appeared in \cite{Re}
and has been discussed in \cite{CDI1,IKK15}.

Consider the algebra $\mathcal{H}$ generated by  $n$ Heisenberg
canonical pairs $ x_a, \dd_a, a= 1, ...,n$,
$$
x_a \dd_b - \e \dd_b x_a =
[x_a, \dd_b]_{-\e} = \e_{ab}, [x_a, x_b]_{-\e} = [\dd_a, \dd_b]_{-\e} =
0,
$$
and the map $\rho: \mathcal{Y}^{(2)}(\mathcal{G}) \to \mathcal{H}$
\be \label{JS}
 \rho(G_{ab}) =  \tilde G_{ab} = x_a\dd_b - \e x_b \dd_a.
\ee
Then we find that $ \tilde G $ is graded antisymmetric and that
the Lie algebra relations are fulfilled.
Note that  the Heisenberg pairs are bosonic in the orthogonal
and fermionic in the symplectic case.

The condition $W_{12} = 0$ holds and by  Proposition 6
all second order evaluation conditions are fulfilled.
The easy way to check the vanishing of $W_{12}$ is to recall the
fact that it is the graded-antisymmetrization of the product
$ G_{a_1b_1}G_{a_2b_2}$ (see Proposition 1)
and use the Heisenberg algebra relations.
It is not difficult to check the RLL relations (\ref{rll}) directly
with $L(u)$ being of the form (\ref{L2G}) and with the substitution (\ref{JS}).

\section{Discussion}

Solutions of the Yang-Baxter relations with orthogonal or symplectic symmetry
and with a simple linear or quadratic dependence on the spectral parameter
exist under restrictions going beyond the Lie algebra relations. In this
paper we have investigated the constraints arising from the truncation of
the expansion in the spectral parameter in general form.

In the linear case, the Yangian algebra $\mathcal{Y}^{(1)}(\mathcal{G})$
is generated by
$\bar G^a_{\ b}$, obeying the underlying Lie algebra as well as
the condition of vanishing of the graded-symmetric traceless part of the
square, $\bar G^2$.

In the quadratic case, the Yangian algebra $ \mathcal{Y}^{(2)}(\mathcal{G})$
is generated
by the matrix elements of $\bar G$ obeying the Lie algebra relations
and the matrix elements of $\bar H$.
The latter transform as the adjoint representation of the Lie algebra.
$\bar G$ and $\bar H$ are related by further conditions.
The commutation relations of the generators contained in the matrix $\bar H$
are given by an expression in terms of the graded anti-symmetrized
product of the generators $\bar G$, $(W_{12})_{abcd} = \bar G_{[ab} \bar
G_{cd)} $.
The anti-commutator of  $\bar G$ and $\bar H$  is expressed by the graded
symmetric part of $\bar G^2$. The square $\bar H^2$ is expressed in terms of
a fourth order polynomial in $\bar G$.

The second order evaluation  $ \mathcal{Y}^{(2)}(\mathcal{G})$
can be further restricted
to the Lie algebra resolution, where $\bar H$  is proportional to  $\bar
G$. Then the constraints are fulfilled by imposing, besides of
relations on the central elements, the single condition of vanishing
of the  graded anti-symmetrized
product of the generators $\bar G$, $(W_{12})_{abcd} = \bar G_{[ab} \bar
G_{cd)} $.

The known example of a linear $L$ operator, representing
$\mathcal{Y}^{(1)}(\mathcal{G})$,
is based on an underlying Clifford (orthogonal case) or oscillator
(symplectic case) algebra. The known example of a quadratic $L$ operator,
representing the Lie algebra resolution of $ \mathcal{Y}^{(2)}(\mathcal{G})$,
is based on the underlying Heisenberg algebra of $n$ canonical pairs,
bosonic in the orthogonal case and fermionic in the symplectic case.

It is instructive to see how the set of constraints is fulfilled by these
constructions in terms of the underlying algebras. This helps to understand
better the distinguished role of these examples.

The general form of the algebra conditions given here allows now to
investigate the set of all representations of the truncated Yangians
 $ \mathcal{Y}^{(1)}(\mathcal{G})$ and
$ \mathcal{Y}^{(2)}(\mathcal{G})$. It is of physical relevance
to see whether there are simple orthogonal or symplectic $R$ operators
essentially different from the ones known so far.

\noindent
{\bf Acknowledgments.}

\noindent
We thank A.P. Isaev for discussions. Our collaboration has  been
supported by JINR (Dubna) via a Heisenberg-Landau grant ( R.K.) and a
Smorodinski-Ter-Antonyan grant ( D.K.). The work of D.K. was partially
supported by the Armenian State Committee of Science grant
 18RF-002 and by Regional Training Network on Theoretical
Physics sponsored by Volkswagenstiftung Contract nr. 86 260.

\vspace{1cm}



\begin{thebibliography}{99}
 \footnotesize\itemsep=0pt


  \bibitem{FST}
        L.D.~Faddeev, E.K.~Sklyanin and L.A.~Takhtajan, {\it The Quantum
                Inverse Problem Method. 1}, Theor.\ Math.\ Phys.\  {\bf 40}
(1980)
        688 [Teor.\ Mat.\ Fiz.\  {\bf 40} (1979) 194]

        \bibitem{TTF}
        V.O.~Tarasov, L.A.~Takhtajan and L.D.~Faddeev, {\it Local
                Hamiltonians for integrable quantum models on a lattice},
Theor.\
        Math.\ Phys.\ {\bf 57} (1983) 163

        \bibitem{KuSk1}
        P.P. Kulish and E.K. Sklyanin ,~{\it Quantum spectral transform
                method. Recent developments}, Lect. Notes in Physics, {\bf v151},
        (1982) , 61-119.

        \bibitem{Fad}
        L.D.~Faddeev,~{\it How Algebraic Bethe Ansatz works for integrable
                model}, In: Quantum Symmetries/Symetries Quantiques,
        Proc.Les-Houches summer school, LXIV. Eds. A.Connes,K.Kawedzki,
        J.Zinn-Justin. North-Holland, 1998, 149-211, [hep-th/9605187]


\bibitem{LevPadua} L. N. Lipatov, ``High-energy asymptotics of muticolor QCD
and exactly solvable lattice models'', JETP Lett. 59(1994) 596,
Padua preprint DFPD/93/TH/70, hep-th/9311037.

\bibitem{FK}
 L.~D.~Faddeev and G.~P.~Korchemsky,
  ``High-energy QCD as a completely integrable model,''
  Phys.\ Lett.\ B {\bf 342} (1995) 311
  doi:10.1016/0370-2693(94)01363-H
  [hep-th/9404173].

\bibitem{BSt}
  N.~Beisert and M.~Staudacher,
  ``The N=4 SYM integrable super spin chain,''
  Nucl.\ Phys.\ B {\bf 670} (2003) 439
  doi:10.1016/j.nuclphysb.2003.08.015
  [hep-th/0307042];

  N. Beisert, Phys. Reports {\bf 405} (2005) 1, hep-th/0407277.


\bibitem{DHP}
  J.~M.~Drummond, J.~M.~Henn and J.~Plefka,
  ``Yangian symmetry of scattering amplitudes in N=4 super Yang-Mills
   theory,''
  JHEP {\bf 0905} (2009) 046
  [arXiv:0902.2987 [hep-th]];



\bibitem{Drinfeld}
V.G. Drinfeld, Hopf algebras and quantum Yang-Baxter
 equations, Sov. Math. Dokl. {\bf 32} (1985) 254-258.

\bibitem{Drin} V.G. Drinfeld,
Quantum Groups, in Proceedings of the Intern. Congress of Mathematics,
Vol. 1 (Berkeley, 1986), p. 798.





        \bibitem{ZZ}
        A.B.~Zamolodchikov and Al.B.~Zamolodchikov,
        ``Factorized S Matrices in Two-Dimensions as the
  Exact Solutions of
        Certain Relativistic Quantum Field Models,''
        Annals Phys.\  {\bf 120} (1979) 253.

        Al.B.~Zamolodchikov,
  "Factorizable Scattering in Assimptotically
 Free 2-dimensional Models of Quantum Field Theory", PhD Thesis, Dubna
(1979), unpublished


        \bibitem{BKWK}
        B.~Berg, M.~Karowski, P.~Weisz and V.~Kurak,
        ``Factorized U(n) Symmetric s Matrices in Two-Dimensions,''
        Nucl.\ Phys.\ B {\bf 134} (1978) 125.
        \bibitem{SW}
        R.~Shankar and E.~Witten,
        ``The S Matrix of the Kinks of the ($\psi^-$bar $\psi$)**2 Model,''
        Nucl.\ Phys.\ B {\bf 141} (1978) 349
        [Erratum-ibid.\ B {\bf 148} (1979) 538].




        \bibitem{Re}
        N.Yu. Reshetikhin, Integrable models of quantum one-dimensional
        models with $O(n)$ and $Sp(2k)$ symmetry,
    Theor. Math. Fiz. {\bf 63}  (1985) 347-366.

        \bibitem{API} A.P. Isaev, Quantum groups and Yang-Baxter equations,
 preprint MPIM (Bonn), MPI 2004-132, \\
 (http://webdoc.sub.gwdg.de/ebook/serien/e/mpi\_mathematik/2004/132.pdf).

 \bibitem{FRT}  L.D. Faddeev,  N.Yu. Reshetikhin and L.A. Takhtajan,
Quantization of Lie groups and  Lie algebras, (Russian)
Algebra i Analiz
{\bf 1} (1989) no. 1, 178--206. English translation in: Leningrad Math. J.
{\bf 1} (1990) no. 1, 193--225.






        \bibitem{CDI1}
        D.~Chicherin, S.~Derkachov and A.~P.~Isaev,
        ``Conformal group: R-matrix and star-triangle relation,''
       JHEP {\bf 04}(2013), 020
        arXiv:1206.4150 [math-ph].

        ``The spinorial R-matrix,''
        J.\ Phys.\ A {\bf 46} (2013) 485201.
        arXiv:1303.4929 [math-ph].


\bibitem{IKK15}
A.~P.~Isaev, D. Karakhanyan, R. Kirschner,
``Orthogonal and symplectic Yangians and Yang-Baxter R operators,''
 Nucl. Phys. {\bf B904} (2016) 124147,
arXiv:1511.06152


\bibitem{FIKK16}
  J.~Fuksa, A.~P.~Isaev, D.~Karakhanyan and R.~Kirschner,
  ``Yangians and Yang-Baxter R-operators for ortho-symplectic superalgebras,''
  Nucl.\ Phys.\ B {\bf 917} (2017) 44
  doi:10.1016/j.nuclphysb.2017.01.029
  [arXiv:1612.04713 [math-ph]].


\bibitem{KK16}
  D. Karakhanyan  and R. Kirschner,
  `` Yang-Baxter relations with orthogonal or symplectic symmetry,''
   J.\ Phys.\ Conf.\ Ser.  {\bf 670} (2016)  no 1,   012029
  doi:10.1088/1742-6596/670/1/012029

\bibitem{AACFR} D. Arnaudon, J. Avan, N. Crampe, L. Frappat, E. Ragoucy,
"R-matrix presentation for
super-Yangians Y (osp(m|2n))", J. Math. Phys. 44 (2003), no. 1, 302-308. arXiv:math/0111325.

\bibitem{AMR} D. Arnaudon, A. Molev, E. Ragoucy,
"On the R-matrix realization of Yangians and their
representations", Ann. Henri Poincar\'e 7 (2006), no. 7-8, 1269-1325. arXiv:math/0511481.

\bibitem{JLM17}
N. Jing, M. Liu, A. Molev,
``Isomorphism between the R-Matrix and Drinfeld presentations of Yangian in
types B,C and D,''
arXiv:1705.08166


\bibitem{GRW} N. Guay, V. Regelskis, C. Wendlandt,
"Equivalences between three presentations of orthogonal
and symplectic Yangians", arXiv:1706.05176.



\end{thebibliography}
\end{document}